\documentclass[aps,prd,final, nofootinbib,superscriptaddress]{revtex4}
\usepackage{amsmath, amsfonts, amsthm, amssymb, graphicx,epsfig,bm,mathrsfs}
\usepackage{bbold}
\usepackage{ytableau}
\theoremstyle{plain}

\theoremstyle{definition}

\def\be{\begin{equation}}
\def\ee{\end{equation}}
\def\ba{\begin{eqnarray}}
\def\ea{\end{eqnarray}}

\def\bdm{\begin{displaymath}}
\def\edm{\end{displaymath}}

\def\bq{\begin{quote}}
\def\eq{\end{quote}}

\def\d{{\rm d}}
\def\del{\partial}
\def\ltap{\ \raise.3ex\hbox{$<$\kern-.75em\lower1ex\hbox{$\sim$}}\ }
\def\gtap{\ \raise.3ex\hbox{$>$\kern-.75em\lower1ex\hbox{$\sim$}}\ }
\def\gl{\ \raise.5ex\hbox{$>$}\kern-.8em\lower.5ex\hbox{$<$}\ }
\def\roughly#1{\raise.3ex\hbox{$#1$\kern-.75em\lower1ex\hbox{$\sim$}}}

 at 11truept

\def \plethysm {\sym^m(\bigwedge^p) \otimes \sym^{m-1}(\sym^2)}
\def \plethysmpart {\sym^4(\bigwedge^3) \otimes \sym^{3}(\sym^2)}

\def\d2{{(\hat D \pi)^2 }}

\def \E {{\mathcal E}}
\def \A {{\mathcal A}}
\def \ap {A}
\def \bp {B}
\def \G {{\mathcal G}}
\def \S {{\mathcal S}}
\def \B {{\mathcal C}}
\def \F {{\mathcal F}}
\def \T {{\mathcal T}}

\def \X {{\mathcal X}}
\def \Y {{\mathcal Y}}

\def \Vspace {{\bm{V}}}
\def \Uspace {{\bm{U}}}
\def \Wspace {{\bm{W}}}

\def \VspaceR {{\bm{V}}_\mathbb{R}}
\def \VspaceC {{\bm{V}}_\mathbb{C}}
\def \Uset {\bm{u}}
\def \Vset {\bm{v}}

\def \STl {ST_\lambda}
\def \Rgroup {R_{\lambda_k}}
\def \Cgroup {C_{\lambda_k}}
\def \Rsym {\bm{r}_{\lambda_k}}
\def \Csym {\bm{c}_{\lambda_k}}
\def \Youngsym {\bm{y}^{sym}_{\lambda_k}}
\def \Younganti {\bm{y}^{anti}_{\lambda_k}}

\def \Youngsymanti {\bm{y}^{sym/anti}_{\lambda_k}}
\def \Youngantisym {\bm{y}^{anti/sym}_{\lambda_k}}
\def \Youngsymantimu {\bm{y}^{sym/anti}_{\mu}}

\def \Youngsymantitilde {\bm{\tilde{y}}^{sym/anti}_{\lambda_k}}
\def \Symop {{\rm Sym}}
\def \Antisymop {{\Lambda}}

\def \bvect {\bm{\mathcal{B}}}
\def \bcovect {\bm{\tilde{\mathcal B}}}
\def \vcovect {\bm{\tilde{\mathcal V}}}
\def \gl {D}

\newcommand{\sign}{\operatorname{sign}}
\newcommand{\sym}{\operatorname{Sym}}
\newcommand{\spann}{\operatorname{span}}

\newcommand{\Tn}{\bm{V^{\otimes n}}}

\newcommand{\beq}{\begin{equation}}
\newcommand{\eeq}{\end{equation}}
\newcommand{\bea}{\begin{eqnarray}}
\newcommand{\eea}{\end{eqnarray}}
\newcommand{\beqa}{\begin{eqnarray}}
\newcommand{\eeqa}{\end{eqnarray}}
\newcommand{\nn}{\nonumber\\}

\def \1 {\pi_{i_1}}
\def \2 {\pi_{i_2}}
\def \3 {\pi_{i_3}}
\def \4 {\pi_{i_4}}
\def \5 {\pi_{i_5}}
\def \6 {\pi_{i_6}}
\def \7 {\pi_{i_7}}

\begin{document}
\hspace{5.2in} \mbox{YITP-15-122, IPMU15-0221}\\

\title{On p-form theories with gauge invariant second order field equations.}
\author{C\'edric Deffayet}
\affiliation{UPMC-CNRS, UMR7095, Institut d'Astrophysique de Paris, GReCO, 98bis boulevard Arago, F-75014 Paris, France}
\affiliation{IHES, Le Bois-Marie, 35 route de Chartres, F-91440 Bures-sur-Yvette, France}
\author{Shinji Mukohyama}
\affiliation{Yukawa Institute for Theoretical Physics, Kyoto University, 606-8502,
Kyoto, Japan}
\affiliation{Kavli Institute for the Physics and Mathematics of the Universe, Todai Institutes for Advanced Study, University of Tokyo (WPI), 5-1-5 Kashiwanoha, Kashiwa, Chiba 277-8583, Japan}
\author{Vishagan Sivanesan}
\affiliation{UPMC-CNRS, UMR7095, Institut d'Astrophysique de Paris, GReCO, 98bis boulevard Arago, F-75014 Paris, France}
\ytableausetup 
{mathmode, boxsize=0.5cm, centertableaux}
\begin{abstract}
We explore field theories of a single p-form with equations of motions of order strictly equal to two and gauge invariance. We give a general method for the classification of such theories which are extensions to the p-forms of the Galileon models for scalars. Our classification scheme allows to compute an upper bound on the number of different such theories depending on p and on the space-time dimension. We are also able to build a non trivial Galileon like theory for a 3-form with gauge invariance and an action which is polynomial into the derivatives of the form. This theory has gauge invariant field equations but an action which is not, like a Chern-Simons theory. Hence the recently discovered no-go theorem stating that there are no non trivial gauge invariant vector Galileons (which we are also able here to confirm with our method) does not extend to other odd p cases. 
\end{abstract}
\maketitle

\section{Introduction}
There has recently been a lot of interest in building and studying scalar theories on flat space-times which have second order field equations non linearly depending on the field and its first and second derivatives (see \cite{Deffayet:2011gz} for the complete construction and classification of these theories in arbitrary dimensions and e.g. \cite{Deffayet:2013lga} for a review of the formal aspects of these theories). 
The interest for such theories, which are for a scalar what Lovelock theories are for a metric (and are in fact known for a long time at least in 4 dimensions \cite{Horndeski:1974wa, Fairlie:1992nb,Fairlie:1992he,Fairlie:1991qe}), has been renewed by the discussions around the so-called Galileons \cite{Nicolis:2008in}: scalar fields on flat space-times with equations of motions only depending on second derivatives . Such theories can be formulated conveniently \cite{Deffayet:2009mn} using the following tensor $\bm{\epsilon_{2(2m)}}$ defined by 
\ba \label{DEFAten}
\epsilon_{2(2m)}^{a_{\vphantom{()}1} a_{\vphantom{()}2}
\ldots a_{\vphantom{()}m} b_{\vphantom{()}1} b_{\vphantom{()}2}
\ldots b_{\vphantom{()}m}} \equiv
\frac{1}{(D-m)!}\,
\epsilon^{a_{\vphantom{()}1}
a_{\vphantom{()}2}  \ldots
a_{\vphantom{()}m} c_{\vphantom{()}1}c_{\vphantom{()}2}\ldots
c_{\vphantom{()}D-m}}_{\vphantom{a_{\vphantom{()}1}}}
\,\epsilon^{b_{\vphantom{()}1} b_{\vphantom{()}2} \ldots
b_{\vphantom{()}m}}_{\hphantom{b_{\vphantom{()}1}
b_{\vphantom{()}2} \ldots
b_{\vphantom{()}2m}}c_{\vphantom{()}1}
c_{\vphantom{()}2}\ldots c_{\vphantom{()}D-m}} \,
\ea
where the totally antisymmetric Levi-Civita tensor is given (on flat space time)
\ba \label{DEFLC}
\epsilon^{a_{\vphantom{()}1} a_{\vphantom{()}2} \ldots
a_{\vphantom{()}D}} \equiv 
\delta^{[a_{\vphantom{()}1}}_1 \delta^{a_{\vphantom{()}2}}_2
\ldots \delta^{a_{\vphantom{()}D}]}_D 
\ea
For future reference, we also stress here that we will denote $\bm{\epsilon_{2}}$ as just  $\bm{\epsilon_{2(2D)}}$ where $D$ is the space time dimension.
Then, the Lagrangians for a scalar Galileon $\pi$ can just be taken to be proportional to 
\ba \label{Lgalpi}
\epsilon_2^{a_{\vphantom{()}1} 
\ldots a_{\vphantom{()}m} b_{\vphantom{()}1}\ldots b_{\vphantom{()}m}} \left(\partial_{a_1} \pi_{b_2}\right)\left( \partial_{b_1} \pi_{a_2} \right)\ldots \left(\partial_{a_{m-1}} \pi_{b_{m-1}} \right)\left(\partial_{a_{m}} \pi\right)\left(\partial_{b_{m}} \pi\right)
\ea
where $\pi_a \equiv \partial_a \pi$, such that one has in particular $\partial_a \pi_b = \partial_b \pi_a$, but also the two key (but trivial for a scalar) properties
\ba \label{identitypi}
\partial_{a_i} \partial_{[b_j} \pi_{b_k]} &=& 0\\ 
\partial_{[a_i}\partial_{a_j]} \pi_{b_k} &=& 0
\ea
 where brackets means antisymmetrization. These properties first lead very simply to the conclusion that the field equations derived from the Lagrangians (\ref{Lgalpi}) only contain second derivatives: indeed after varying one term proportional to derivative(s) of $\pi$ in (\ref{Lgalpi}) and integrating by part, the only possible way to distribute the derivative(s) acting on this term is to let it (them) hit another $\pi$ differentiated only once, all terms containing more than two derivatives after this distribution will vanish as a consequence of the identities above where the antisymmetrization comes from the contraction with the tensor $\bm{\epsilon_2}$. These identities also lead to an easy generalization to p-forms (single or multifield) explained in \cite{Deffayet:2010zh} (see also \cite{Padilla:2010de,Padilla:2010tj,Hinterbichler:2010xn,Trodden:2011xh,Sivanesan:2013tba,Padilla:2012dx} for subsequent works on the multi-scalar case). Indeed, e.g. for a single p-form ${\bm\A}$ one just need to replace in (\ref{Lgalpi}) $\pi_{a}$ by some p-form field strength $\bm{\F} = d \bm{\A}$ with components $F_A$ (where $A$ now means as set of $p+1$ indices) and in the last two terms in (\ref{Lgalpi}) by the first derivatives the the form. E.g. for a 3 form $\bm{\A}$ we would get the action (needing 9 space-time dimensions) 
\be \label{gefdef3}
S = \int_{\cal M} d^9 x \epsilon^{a_1 \dots a_9} \epsilon^{b_1 \ldots b_9} \left(\partial_{a_1}\F_{B_1} \right)\left(\partial_{b_1} \F_{A_1}\right) \left( \partial_{a_6} \A_{B_2} \right) \left(\partial_{b_6} \A_{A_2}  \right)
\ee
where here $A_1 =\{a_2,a_3,a_4,a_5\}$, $A_2 = \{a_7,a_8,a_9\}$ $B_1= \{b_2,b_3,b_4,b_5\}$ and $B_2 = \{b_7,b_8,b_9\}$,  $\A$ is by assumption a 3 form and $\F = d \A$ is the associated Field strength. 
The role of the first identity in (\ref{identitypi}) is here played by the Bianchi identity for the field strength $\partial_{[b} \F_{B]} = 0,$ while the equivalent of the second identity still holds as it is just a consequence of the commuting of partial derivatives on flat space-times. The above construction, spelled out in Ref. \cite{Deffayet:2010zh}, allows to get non trivial theories for single even-p-forms and multi p-forms retaining also gauge invariance, however it fails for single odd p forms in particular. Indeed, e.g. the above action 
(\ref{gefdef3}) has vanishing field equations. 

 In Ref. \cite{no-go} it was in fact proven that no single vector Galileon could be constructed.\footnote{Note however that non trivial vector Galileons can be obtain if one relaxes 
the hypothesis of gauge invariance \cite{Heisenberg:2014rta,Hull:2014bga,Hull:2015uwa,Tasinato:2014eka,Khosravi:2014mua,Charmchi:2015ggf,Allys:2015sht}} 
 I.e. under the assumption that the theory had gauge invariance, had an action principle and had field equations depending only on derivatives of order less or equal to two, it was shown that one could only obtain field equations linear in second derivatives. The main purpose of this paper is to investigate the same issue for other odd p (i.e. p odd and larger than 1). In this way we will provide a method that gives an upper bound on the number of different Galileon like theories of single p-forms (and p of any parity). This will in particular provide a new proof of the results of \cite{no-go} but also allows us to construct a non trivial Galileon like theory for a 3-form. 

The paper is organized as follows. In the following section, we derive necessary conditions in order for a single p-form theory to have gauge invariant field equations containing derivatives of order less or equal to 2 and to have an action principle. These conditions are expressed as symmetry conditions on the field equations as well as on derivatives of the field equations. We then (section \ref{section3}) introduce some tools we use later to analyze these symmetries. In the following section \ref{section4} we derive an upper bound on the number of allowed theories fulfilling our criteria. In the last section (section \ref{section5}) we show that our formalism allows one to give a simple proof of the no-go theorem stated in Ref.\cite{no-go} for vector Galileons and construct an example of a non trivial 3-form Galileon like theory. We then conclude in section \ref{section6}. We have gathered in the appendices various abstract (some standard and some less known) results used in the course of our work in order to make it self contained.  

\section{Necessary conditions for the existence of a non trivial Galileon p-form}
\subsection{Derivation}
Our goal here is to first derive a set of necessary conditions for the existence of a Galileon p-form theory. I.e. we look for a theory of a p-form $\bm{\A} \in \bm{\bigwedge^p}$ of components $\A_{a[p]}$ (denoting by $\bm{\bigwedge^p}$ the set of p-forms), such that the field equations of this p-form do not contain derivatives of order higher than two.
Note that here and in the following $a[p]$ denotes an ordered set of $p$ indices $a_1,a_2,\cdots a_p$ carried by some object which is antisymmetric in all the indices in the string of indices $a[p]$, i.e., for the case at hand, such that for any permutation $\sigma$ belonging to the permutation group $S_p$ of $p$ objects, we have $\A_{a_{\sigma(1)}a_{\sigma(2)}\cdots a_{\sigma(p)}} = \epsilon(\sigma) \A_{a_1 a_2 \cdots a_p}$ where $\epsilon(\sigma)$ is the signature of the permutation $\sigma$. 
Similarly we will denote simply by $a\{p\} = \{a_1 \dots a_p\}$ an ordered string of indices carried by some object, and by $a(p)$ such a string assuming in addition that the object $\A$ which carries this string is symmetric into the corresponding indices, i.e. such that for every permutation $\sigma$ of the symmetric group $S_p$ we have 
$\A_{a_{\sigma(1)}a_{\sigma(2)}\cdots a_{\sigma(p)}} = \A_{a_1 a_2 \cdots a_p}\equiv \A_{a(p)}$. Sometimes, we will have to pull out a given index out of such strings of indices and we will denote this operation by the following notation:
$a[p] \equiv a[p-1]a_p$ meaning that the objects which carries the string $a[p-1]a_p$ is antisymmetric on all indices including $a_p$, but also that the order of the first $p-1$ indices is the same in $a[p-1]$ and $a[p]$. Furthermore, we will always denote by lower case latin letters space-time indices and, in order to alleviate notations, with an upper case latin letter a string 
of "antisymmetric" indices such as $a[p] \equiv \ap$ or $b[p]\equiv \bp$, and furthermore we will then use the same notation to denote a string of $p$ indices or of a different length, (i.e. $\bp$ can denotes e.g. a string such as $b[p]$ or $b[p-1]$) whenever there is no risk of ambiguity.\footnote{For example a p-form $\A$ will always carry p "antisymmetric" indices and hence its component will sometimes be denoted as $\A_\ap$ or $\A_\bp$ when there is no ambiguity, or equivalently $\A_{a[p]}$ or $\A_{b[p]}$ when there are.}
We will also use the following notations to denotes derivatives of a quantity such as $\E^{a[p]}\equiv \E^\ap$ with respects to $\A_{b[p]}\equiv \A_\bp$ or its successive partial derivatives\footnote{these derivatives being noted as usual as $\A_{A,b}, \A_{A,bc}, \dots$} (always denoted by a {\it{comma}})
\begin{eqnarray} \E^{\ap|\bp} &\equiv& \E^{a[p]|b[b]}   \equiv \frac{\del \E^{a[p]}}{\del \A_{b[p]}} \equiv \frac{\del \E^{\ap}}{\del \A_{\bp}} , \\
\E^{\ap|\bp,c} &\equiv& \E^{a[p]|b[p],c} \equiv \frac{\del \E^{a[p]}}{\del \left(\del_c \A_{b[p]}\right)}\equiv \frac{\del \E^{a[p]}}{\del \A_{b[p],c}} \equiv \frac{\del \E^{\ap}}{\del \A_{\bp,c}},\\
\E^{\ap|\bp,cd} &\equiv& 
\E^{a[p]|b[p],cd} \equiv  \frac{\del \E^{a[p]}}{\del \left(\del_c \del_d \A_{b[p]}\right)} \equiv \frac{\del \E^{a[p]}}{\del \A_{b[p],cd}}\equiv \frac{\del \E^{\ap}}{\del \A_{\bp,cd}}.
\end{eqnarray}
Having introduced these notations, we first adapt the derivation of \cite{no-go} (valid for a 1-form) to the case of an arbitrary p-form with an action 
\be 
\S = \int d^D x \,\, {\cal L}[\A_{\bp}; \del_a \A_{\bp}; \del_a \dots \del_b \A_{\bp}],
\ee
yielding the equations of motion
\be \label{Field0}
\E^{\ap} \equiv \frac{\delta \S }{\delta \A_{\ap}}   = 0.
\ee
We demand that these equations do not contain derivatives of order higher than two, i.e. 
\be 
\E^{\ap} = \E^{\ap}(\A_{\bp};\A_{{\bp},a};\A_{{\bp},ab}).  \label{Cond0}
\ee
The fact that $\E^{\ap}$ derives from an action principle gives non trivial integrability conditions that we now derive.
We first use the commutativity of the functional derivatives
\be \label{comm}
\left [\frac{\delta}{\delta \A_{\bp}(y)},\frac{\delta}{\delta \A_{\ap}(x)} \right] \S = 0,
\ee
which we can rewrite using 
\be 
\frac{\delta}{\delta \A_{\bp}(y)}\frac{\delta}{\delta \A_{\ap}(x)} \S \equiv \frac{\delta  \E^{\ap}(x)}{\delta \A_{\bp}(y)}.
\ee
Here
\be 
\delta \E^{\ap}(x) = \int d^D y' \delta \A_{\bp}(y')\left\{ \delta(x-y') \E^{{\ap}|{\bp}}(y') - \left(\delta(x-y')\E^{{\ap}|{\bp},c}\right)_{,c} + \left( \delta(x-y')\E^{{\ap}|{\bp},cd} \right)_{,cd} \right\},
\ee
such that 
\be 
\frac{\delta \E^{\ap}(x)}{\delta \A_{\bp}(y)} = \delta (x-y)\E^{{\ap}|{\bp}}(y) -\left(\delta(x-y)\E^{{\ap}|{\bp},c}\right)_{,c} + \left(\delta(x-y)\E^{{\ap}|{\bp},cd} \right)_{,cd}.
\ee
Using a test function $\G(x)$ to integrate over \eqref{comm} and subsequently integrating by parts we get 
\ba 
0 &=& \int d^D y \; \G(y)\left\{ \left [\frac{\delta}{\delta \A_{\bp}(y)},\frac{\delta}{\delta \A_{\ap}(x)} \right] \S\right\}\nn
&=& \G \left\{ \E^{{\ap}|{\bp}}-\E^{{\bp}|{\ap}} + \left( \E^{{\bp}|{\ap},c}\right)_{,c} -\left(\E^{{\bp}|{\ap},cd} \right)_{,cd}\right\}\nn
&&+\G_{,c} \left \{ \E^{{\ap}|{\bp},c}+\E^{{\bp}|{\ap},c} -2 \del_d \E^{{\bp}|{\ap},cd}\right \} + \G_{,cd} \left\{ \E^{{\ap}|{\bp},cd}-\E^{{\bp}|{\ap},cd}\right\},
\ea
where in the second equality above all the arguments are evaluated at the same space time point $x$. 
Since $\G(x)$ is arbitrary we get the integrability conditions,
\ba  
&& \E^{{\ap}|{\bp}}-\E^{{\bp}|{\ap}}+ \left( \E^{{\bp}|{\ap},c}\right)_{,c} -\left(\E^{{\bp}|{\ap},cd} \right)_{,cd} =0, \label{IC1}\\
&& \E^{{\ap}|{\bp},c}+\E^{{\bp}|{\ap},c} -2 \del_d \;\E^{{\bp}|{\ap},cd} =0, \label{IC2}\\
&& \E^{{\ap}|{\bp},cd}-\E^{{\bp}|{\ap},cd} =0. \label{IC3}
\ea

By taking derivative of \eqref{IC2} with respect to $\A_{C,efg}$, we also get
\be \label{IC4}
\E^{\ap| B_1,{\overset{\frown}{c_1}} d_1|B_2,{\overset{\frown}{c_2}} {\overset{\frown}{d_2}}} =0=  \E^{\ap| B_1,c_1 {\overset{\frown}{d_1}}|B_2,{\overset{\frown}{c_2}} {\overset{\frown}{d_2}}},
\ee
where here and henceforth a horizontal parenthesis, $\frown$, denotes a symmetrization on the corresponding indices. 
The above constraints (\ref{IC1}-\ref{IC4}) just expresses that the field equations we consider, i.e. (\ref{Field0}) derive from an action. 
Note that (\ref{IC3}) shows that $\E^{{\ap}|{\bp},cd}$ is symmetric under the exchange of the group of indices $A$ and $B$. This extends further to the derivatives of $\E^{{\ap}|{\bp},cd}$ with respect to the second derivatives of the form. 
E.g. consider $\E^{{\ap}|{\bp},cd|{C},ef}$, we have by standard commutations of derivatives 
\be \label{COMMUTEDERIVE}
\E^{{\ap}|{\bp},cd|{C},ef} = \E^{{\ap}|{C},ef|{\bp},cd}
\ee
while (\ref{IC3}) implies obviously that 
\be
\E^{{\ap}|{\bp},cd|{C},ef} = \E^{{\bp}|{\ap},cd|{C},ef}.
\ee
These two identities can be used further to show the symmetry under the exchange of $A$ and $C$ as well as $B$ and $C$ respectively as shown below
\ba
\E^{{\ap}|{\bp},cd|{C},ef} &=& \E^{{\ap}|{C},ef|{\bp},cd} = \E^{{C}|{\ap},ef|{\bp},cd} = \E^{{C}|{\bp},cd|{\ap},ef} \label{SYMACproof}\\
\E^{{\ap}|{\bp},cd|{C},ef} &=& \E^{{\bp}|{\ap},cd|{C},ef} = \E^{{\bp}|{C},ef|{\ap},cd}= \E^{{C}|{\bp},ef|{\ap},cd} =\E^{{C}|{\ap},cd|{\bp},ef}=\E^{{A}|{C},cd|{\bp},ef}. \label{SYMABCproof}
\ea
This further shows (using the last identity above as well as (\ref{COMMUTEDERIVE})) that $\E^{{\ap}|{\bp},cd|{C},ef}$ is also symmetric under commutations of the pair of indices $\{c,d\}$ and $\{e,f\}$, namely that
\ba
\E^{{\ap}|{\bp},cd|{C},ef} &=&\E^{{\ap}|{\bp},ef|{C},cd}.
\ea

We now derive some extra constraints demanding that these field equation shall be gauge invariant.  This implies invariance under,
\be 
\bm{\A} \to \bm{\A} + d\bm{\B} \equiv \bm{\A'} \quad\quad \bm{\B} \in {\bm{\bigwedge^{p-1}}}; \bm{\A}, \bm{\A'}  \in \bm{\bigwedge^{p}}.
\ee
In components, this induces the following transformation on $\bm{\A}$ and it's derivatives,
\ba 
&& \A_{{a[p-1]}a_p} \to \A_{{a[p-1]}a_p} + \B_{[{a[p-1]},a_p]} \equiv \A'_{{a[p-1]}a_p}, \\
&& \A_{{a[p-1]}a_p,c} \to \A_{{a[p-1]}a_p,c} + \B_{[{a[p-1]},a_p]c} \equiv \A'_{{a[p-1]}a_p,c}, \\
&& \A_{a[p-1]a_p,cd} \to \A_{a[p-1]a_p,cd} + \B_{[a[p-1],a_p]cd} \equiv \A'_{a[p-1]a_p,cd}.
\ea
We thus demand that 
\be \label{gaugeinvariance}
\E^A(\A'_{A};\A'_{A,c};\A'_{A,cd}) = \E^A(\A_{A};\A_{A,c};\A_{A,cd}).
\ee
By taking derivative of (\ref{gaugeinvariance}) with respect to $\B_{[a[p-1],a_p]cd}$, $\B_{[{a[p-1]},a_p]c}$ and $\B_{[{a[p-1]},a_p]}$ respectively and setting $\B_{a[p-1]} =0$ we get,
\ba 
&& \E^{\ap|b[p-1](b_{p},cd)} =0, \label{G3}\\
&& \E^{\ap|b[p-1](b_p,c)} =0,  \label{G2}\\
&& \E^{\ap|\bp} =0. \label{G1}
\ea
Using \eqref{IC3} and \eqref{G3} we have,
\ba 
\E^{\ap|b_1 \dots {\overset{\frown}{b_r}} \dots b_p,{\overset{\frown}{c}}{\overset{\frown}{d}}} = 0= \E^{a_1 \dots {\overset{\frown}{a_s}} \dots a_p|\bp,{\overset{\frown}{c}}{\overset{\frown}{d}}} \label{G3.1}.
\ea
We can use this to further show that $\E^{\ap|B,cd}$ vanishes when symmetrized on any three arbitrary space-time indices.
Indeed, the above (\ref{G3.1}) implies obviously that 
\ba 
\E^{a[p-1]{\overset{\frown}{a_p}}| b[p-1]{\overset{\frown}{b_p}},{\overset{\frown}{c}}{\overset{\frown}{d}}}=0, 
\ea
which we can expand as
\ba \label{EXPANDSYM}
\E^{a[p-1] a_p | b[p-1]{\overset{\frown}{b_p}},{\overset{\frown}{c}}{\overset{\frown}{d}}}+ \E^{a[p-1]{\overset{\frown}{b_p}}| b[p-1]a_p,{\overset{\frown}{c}}{\overset{\frown}{d}}}+ \E^{a[p-1]{\overset{\frown}{c}}| b[p-1]{\overset{\frown}{b_p}},a_p{\overset{\frown}{d}}}+ \E^{a[p-1]{\overset{\frown}{d}}| b[p-1]{\overset{\frown}{b_p}},{\overset{\frown}{c}} a_p}=0. 
\ea
The first two terms above vanish as a consequence of (\ref{G3.1}), while the trivial 
\ba
\E^{A|B,cd} = \E^{A|B,dc}
\ea
implies the equality of the last two terms on the left hand side of (\ref{EXPANDSYM}), yielding 
\ba \label{NEWRELA1}
\E^{a[p-1]{\overset{\frown}{a_p}}| b[p-1]{\overset{\frown}{b_p}},{\overset{\frown}{c}}d}=0.
\ea
The above proven vanishing upon symmetrization of any three indices will play a key role in the following and in fact extends to the derivatives of $\E^{\ap|B,cd}$ with respect to the second derivative of the form, as we now show. We have already obtained the relations (\ref{IC4}). Now consider the relation 
\ba
\E^{a[p-1]{\overset{\frown}{a_p}}| b[p-1]{\overset{\frown}{b_p}},cd| c[p-1]{\overset{\frown}{c_p}},{\overset{\frown}{e}}f} =0 
\ea
obtained from (\ref{NEWRELA1}) and (\ref{COMMUTEDERIVE}). 
Expanding this relation as 
\ba
0&=&\E^{a[p-1]a_p| b[p-1]{\overset{\frown}{b_p}},cd| c[p-1]{\overset{\frown}{c_p}},{\overset{\frown}{e}}f} + \E^{a[p-1]{\overset{\frown}{b_p}}| b[p-1]a_p,cd| c[p-1]{\overset{\frown}{c_p}},{\overset{\frown}{e}}f}\nonumber\\
&& + \E^{a[p-1]{\overset{\frown}{c_p}}| b[p-1]{\overset{\frown}{b_p}},cd| c[p-1]a_p,{\overset{\frown}{e}}f} + \E^{a[p-1]{\overset{\frown}{e}}| b[p-1]{\overset{\frown}{b_p}},cd| c[p-1]{\overset{\frown}{c_p}},a_p f}, 
\ea
and noticing that the first three terms vanish by virtue of (\ref{NEWRELA1}) and (\ref{IC3}), we get that 
\ba
\E^{a[p-1]{\overset{\frown}{a_p}}| b[p-1]{\overset{\frown}{b_p}},c d| c[p-1]{\overset{\frown}{c_p}},e f} =0.
\ea
From (\ref{G3.1}) and (\ref{COMMUTEDERIVE}), we also have 
\ba
\E^{a[p-1]{\overset{\frown}{a_p}}| b[p-1]{\overset{\frown}{b_{p}}},cd| c[p-1]c_p,{\overset{\frown}{e}} {\overset{\frown}{f}}} =0,
\ea
which yields, using the same expansion technique as above,
\ba
\E^{a[p-1]{\overset{\frown}{a_p}}| b[p-1]{\overset{\frown}{b_p}},c d| c[p-1]c_p,{\overset{\frown}{e}} f} =0.
\ea 
Similarly, from (\ref{IC4}), we have 
\ba
\E^{a[p-1]{\overset{\frown}{a_p}}| b[p-1]b_p,{\overset{\frown}{c}}d| c[p-1]c_p,{\overset{\frown}{e}} {\overset{\frown}{f}}} =0,
\ea
which yields, using the same expansion technique as above,
\ba
\E^{a[p-1]{\overset{\frown}{a_p}}| b[p-1]b_p,{\overset{\frown}{c}} d| c[p-1]c_p,{\overset{\frown}{e}} f} =0.
\ea 
A last property of importance can be obtained from the following relation which follows from (\ref{IC4}) yielding
\ba
\E^{A|B_1,{\overset{\frown}{c_1}} {\overset{\frown}{d_1}}|B_2,{\overset{\frown}{c_2}} d_2|B_3,{\overset{\frown}{c_3}} d_3} =0 
\ea
which, after expanding, gives 
\ba
\E^{A|B_1,{\overset{\frown}{c_1}} d_1|B_2,{\overset{\frown}{c_2}} d_2|B_3,{\overset{\frown}{c_3}} d_3} =0 
\ea

In the following, we will denote as $\bm{\E^m}$ the tensor obtained by differentiating the field equation operator $\bm{\E}$ with respect to the second derivatives of the p-form $m-1$ times. I.e., one has by definition $\bm{\E^1}\equiv \bm{\E}$ and in full generality $\bm{\E^m}$ is a rank $(pm + 2(m-1))$ contravariant tensor whose components given by 
\be \label{defEm}
\left(\bm{\E^m}\right)^{\ap \bp_1 c_1d_1 \dots \bp_{m-1} c_{m-1}d_{m-1}}= \E^{\ap|\bp_1,c_1d_1|\dots |\bp_{m-1},c_{m-1}d_{m-1}} \equiv \frac{\del^{m-1} \E^{\ap}}{\del \A_{\bp_1,c_1d_1} \dots \del \A_{\bp_{m-1},c_{m-1}d_{m-1}}}. 
\ee
As will be summarized in the next subsection, the relations derived above for $m=2, 3$ and $4$ extend in fact to arbitray values of $m$, and this will be used below to show that, for a given space-time dimension $D$ (and a given $p$),  $\bm{\E^m}$ vanishes for a large enough value of $m$. This means that under the hypotheses spelled out above, the field equations have to be polynomial in the second derivatives. We will also further restrict the possible theories fulfilling these hypotheses. Introducing here the integer $m_{max}$ such that $\bm{\E^{m_{\max}+1}}$ vanishes but $\bm{\E^{m_{\max}}}$ does not, the idea is then to use the method exposed below to construct explicitly non trivial tensors $\bm{\E^{m_{\max}}}$ which do not depend any more on the second derivatives of the p-form. In the following, we will hence assume that $m_{\max}$ (and hence non trivial $m$) is always larger or equal to 2, since otherwise, the field equations do not depend on second derivatives of the p-form at all.

\subsection{Summary of the symmetries to be used} \label{2.B}

As stated at the end of the previous subsection, our strategy in the rest of the present paper is to seek an explicit form of the tensor $\bm{\E^{m_{\max}}}$, which is $m_{\max}-1$ th derivative of the field equation $\E$ with respect to the second derivatives of the p-form. After that, we can integrate it $m_{\max}-1$ times with respect to the second derivatives of the p-form to obtain $\E$. In each step of integration, the integration ``constant'' may in principle depend on first derivatives of the p-form. For simplicity, however, we shall not introduce such dependence on first derivatives since the main purpose of the present paper is to develop a general formalism and just to show an explicit non-trivial example. The integration ``constants'' cannot depend on the p-form itself without derivatives acted on it because of (\ref{G1}). Therefore, among various symmetry conditions derived in the previous section, for our purpose it suffices to concentrate on those that involve only the derivatives of $\E$ with respect to the second derivatives of the p-form and the starting point is the series of relations derived in the previous subsection from the conditions on the field equations in order for them to derive from an action (\ref{IC1}-\ref{IC4}), be of order at most two in the derivatives (\ref{Cond0}) and be gauge invariant (\ref{G1}-\ref{G3}). From those relations it is easy to derive the corresponding relations satisfied by the tensor $\bm{\E^m}$. It then turns out to be convenient to divide these symmetry relations possessed by the tensor $\bm{\E^m}$ into three categories defined below. We shall use the notation of equation (\ref{defEm}) for the names and the order of the indices of $\bm{\E^m}$.  
\begin{enumerate}
\item \label{set1}
The first set consists of
\begin{enumerate}
\item Antisymmetry within each group of p indices, $\ap$ and $\bp_i$ 
\item Invariance of the tensor under the p-wise interchange of the group of indices: $\bp_i \leftrightarrow \bp_j, \quad i \neq j \quad i,j \in \{1,\dots, (m-1)\}$,
as well as under the exchange between $A$ and any $B_i$. 
\end{enumerate} 
\item \label{set2}
The second set of symmetries consists of
\begin{enumerate}
\item Symmetry of each pair of indices $(c_i, d_i)$
\item Invariance under the pairwise interchange of the indices : $(c_i,d_i) \leftrightarrow (c_j,d_j), i\neq j \quad i,j \in \{1,\dots , (m-1)\}$
\end{enumerate}
\item \label{set3}
The last set of symmetries is the condition that symmetrizing over any 3 indices in $\bm{\E^m}$ yields a vanishing tensor.
\end{enumerate}

Our goal here is to use these symmetries in order to further characterize the possible tensors $\bm{\E^m}$ and eventually construct some explicit example. To do so, we will use the theory of representations of the symmetric group and its link to tensor symmetries, some elements about these issues are given below as well as in appendices.

\section{Tensor symmetries and plethysms} \label{section3}

\subsection{Generalities about tensor symmetries, symmetric and antisymmetric tensors} 
 
Here, we are mainly interested in characterizing the symmetries of the tensor $\bm{\E^m}$ and hence will only discuss the case of contravariant tensors. We denote by $\Vspace$ the tangent vector space of spacetime covectors at a given point and stress again that we will work here only on flat space-time. We denote by $\Tn = \Vspace\otimes \Vspace \otimes \cdots \otimes \Vspace$ the vector space (over the field $\mathbb{R}$ of real numbers) of rank $n$ (contravariant) tensors considered as the n-th tensor product of the vector space $\Vspace$. The vector space $\Tn$ has simple vector subspaces given respectively by the set of totally symmetric rank $n$ tensors that we denote as 
$\sym^n$ and the set of totally antisymmetric rank $n$ tensors that we denote as $\bigwedge^n$. We also define as 
 $\{\bvect_1, \dots , \bvect_D\}$ (where $D$ is the dimension of space-time) base vectors of the vector space $\Vspace$, and  $\{\bcovect^1, \dots , \bcovect^D\}$ the dual basis of covectors (one-forms) such that 
$\bcovect^c\left(\bvect_d\right) = \delta^c_d$. Then, the $n$-tensor products of the $\bvect$ form a basis of $\Tn$. To characterize symmetries of tensors in general, it is useful to use the fact that the permutation group $S_n$ of a set of n distinct elements acts in a natural way on a tensor $\bm{\T}$ of 
components $\T^{a_1 \dots a_n}$, as, given a permutation $\sigma$ inside $S_n$,
\ba \label{actiontensor}
\sigma: \bm{\T}\equiv \T^{a_1 \dots a_n}\, \bvect_{a_1} \otimes \dots \otimes \bvect_{a_n} &\to & \T^{a_1 \dots a_n}\, \bvect_{a_{\sigma^{-1}(1)}} \otimes \dots \otimes \bvect_{a_{\sigma^{-1}(n)}}\nn
&=& \T^{a_{\sigma(1)} \dots a_{\sigma(n)}} \bvect_{a_1} \otimes \dots \otimes \bvect_{a_n} \nn
&\equiv& \sigma(\bm{\T}),
\ea 
where Einstein summation convention is implied. 
I.e. for a given tensor $\bm{\T}$ of components $\T^{a_1 \dots a_n}$, and a given permutation $\sigma$ of $S_n$, $\sigma(\bm{\T})$ is the tensor of components $\left[\sigma\left(\T\right)\right]^{a_1 \dots a_n} = \T^{a_{\sigma(1)} \dots a_{\sigma(n)}}$. The tensors $\sigma(\bm{\T})$ are sometimes called the isomers of $\bm{\T}$  (see e.g. \cite{FiedlerHab}). Note that this action is an action on the places of indices and not on their values; e.g. for a 3-tensor $\T^{a_1 a_2 a_3}$ and the 3-cycle $\sigma = (123)$, one has 
$[\sigma\left(\T\right)]^{112} = \T^{121}$, (and not $\T^{223}$). Note further that with such a definition, the composition of two permutations $\rho$ and $\sigma$ obviously acts as follows on a given tensor $\bm{\T}$, $\left[\rho\left(\sigma\left(\T\right) \right)\right]^{a_1 \cdots a_n} = \T^{a_{\rho\left(\sigma(1)\right)} \cdots a_{\rho\left(\sigma(n)\right)}}$, i.e. it acts by left multiplication on the indices labels. 

The linear transformations on $\Tn$ which commute with the above defined action of the symmetric group $S_n$ on tensors play an important role in the following. These transformations are called bisymmetric transformations and defined by the following action on a given tensor $\bm{\T}$ 
\ba \label{actionGL}
 \bm{\T}\equiv \T^{a_1 \dots a_n}\, \bvect_{a_1} \otimes \dots \otimes \bvect_{a_n} \to  \gl^{a_1 \cdots a_n }_{\hphantom{a_1\cdots a_n} b_1 \cdots b_n}  \T^{b_1 \dots b_n} \bvect_{a_1} \otimes \dots \otimes \bvect_{a_n},  
\ea
where the $D^{2n}$ real numbers $\gl^{a_1 \cdots a_n }_{\hphantom{a_1\cdots a_n} b_1 \cdots b_n}$ verify the "bisymmetry" condition 
\ba \label{bisymcond}
\gl^{a_1 \cdots a_n }_{\hphantom{a_1\cdots a_n} b_1 \cdots b_n} = \gl^{a_{\sigma(1)} \cdots a_{\sigma(n)} }_{\hphantom{a_{\sigma(1)}\cdots a_{\sigma(n)}} b_{\sigma(1)} \cdots b_{\sigma(n)}}, 
\ea
for any permutation $\sigma$ belong to $S_n$. Among the bisymmetric transformations, the one given by 
$\gl^{a_1 \cdots a_n }_{\hphantom{a_1\cdots a_n} b_1 \cdots b_n} = \gl^{a_1}_{\hphantom{a_1} b_1} \cdots \gl^{a_n}_{\hphantom{a_n} b_n}$ where $\gl^{a}_{\hphantom{a} b}$ is an invertible matrix (considered here as a one time covariant and one time contravariant tensor) correspond to the natural action of the general linear group $GL_D$ on a tensor $\bm{\T}$. The set of bisymmetric transformations is the largest set of transformation which commute with the transformation of $S_n$ defined as in (\ref{actiontensor}), and conversely, the set of the symmetric group transformations (\ref{actiontensor}) is the maximal set of transformations which commute with all the bisymmetric transformation (see e.g. \cite{Boerner} p 134-136). This property is at the heart of the so-called Schur-Weyl duality which allows to simultaneously reduce the representations of the symmetric group and of the general linear group on the space of tensors $\Tn$.

Tensor symmetries can in general be represented by one or several relations between tensor components, namely one or several relations of the type
\begin{eqnarray} \label{defsym}
 \T^{a_1 a_2 \cdots a_n} &=& \sum_{\sigma \in S_n} k_{\sigma}  \T^{a_{\sigma(1)} \dots a_{\sigma(n)}}, 
\end{eqnarray}
where $k_{\sigma}$ are real numbers indexed by elements $\sigma$ of the symmetric group $S_n$. Relation (\ref{defsym}) obviously reads, using the above defined action of the symmetric group,
$ \bm{\T} = \sum_{\sigma \in S_n} k_{\sigma}  \times \sigma \left(\bm{\T}\right).$
For future reference we also define the vector space (inside $ \Tn$) of symmetric tensors $\sym^n$, as the set of tensors $\bm{\T} \in \Tn$ obeying $\sigma(\bm{\T}) = \bm{\T},$ for any $\sigma \in S_n$, 
\be 
\sym^n =  \spann \{ \bm{\T}\;|\; \sigma(\bm{\T}) = \bm{\T} \quad \text{for any}\, \sigma \in S_n \}.
\ee
Similarly, the set of antisymmetric tensors $\bigwedge^n$ (i.e. $n-$forms) is the vector subspace of $\Tn$ 
\be 
\bigwedge^n = \spann \{\bm{\T} \;|\; \sigma(\bm{\T}) = \sign(\sigma) \bm{\T} \quad \text{for any}\, \sigma \in S_n    \}.
\ee
where $\sign(\sigma)$ denotes the signature of the permutation $\sigma$.
We can define tensor product spaces using these such as,
\be 
\sym^m \otimes \bigwedge^n = \spann \{ \bm{\X} \otimes \bm{\Y} \;|\; \bm{\X} \in \sym^m, \bm{\Y}\in \bigwedge^n   \}
\ee

Note that, as explained in appendix \ref{symclass}, symmetries of tensors as defined in (\ref{defsym}) can be suitably dealt with using the group algebra of the symmetric group $S_n$ and its decompositions into irreducible spaces under the action of $S_n$. The decomposition uses Young diagrams and tableaux which in turn are associated with irreducible components of the so-called symmetry classes under the action of the bisymmetric transformations (\ref{actionGL}). For example the symmetry class of totally symmetric tensors $\sym^n$ is generated (in the sense explained in appendix \ref{symclass}) by the (only) Young symmetrizer generated from the standard tableau with one line and n boxes filled with $\{1,2,\cdots,n\}$. More generally, as explained in the same appendix, one can consider symmetry classes $\Vspace_{\lambda}$ generated by some specific Young symmetrizers corresponding to the tableau $\lambda$.

\subsection{Plethysms} \label{Plethysms}

Another way to combine symmetric or antisymmetric tensors goes as follows. Consider first a set of $k$ tensors $\bm{\X}_i \in \bigwedge^n$ with $i\in \{ 1,\dots ,k\}$. We can define a tensor 
$\bm{\X}$ which is an element in the \textit{composite} subspace denoted $\sym^k(\bigwedge^n)$ by
\be
\bm{\X} \equiv \sum_{\sigma \in S_k} \bm{\X}_{\sigma(1)} \otimes \dots \otimes \bm{\X}_{\sigma(k)}.
\ee
The tensor $\bm{\X}$ can be written in components
\be 
\bm{\X} = \X^{a^1[n] \dots  a^k[n]} \,(  \bvect_{a^1_1} \otimes \dots \otimes \bvect_{a^1_n}) \otimes \dots \otimes ( \bvect_{a^k_1} \otimes \dots \otimes \bvect_{a^k_n} ),
\ee
where $a^i[n] \equiv \{ a^i_1 \dots a^i_n \}\equiv A^i$ are groups of anti-symmetric indices and the following symmetry under the $n$-wise exchange of indices holds,
\be 
\X^{A^{\sigma(1)} \dots  A^{\sigma(k)}} = \X^{A^1 \dots A^k} \quad \forall \sigma \in S_{k}.
\ee
Similarly, considering now a set of $l$ symmetric tensors of rank $m$, $\bm{\Y}_j$ with $j \in \{1,\dots,l\}$, we define an element $\bm{\Y}$ of the space $\bigwedge^k(\sym^m)$ by
\be 
\bm{\Y} = \sum_{\sigma \in S_l} \sign(\sigma) \;\bm{\Y}_{\sigma(1)} \otimes \dots \otimes \bm{\Y}_{\sigma(l)}.
\ee
In components $\bm{\Y}$ can be expressed as 
\be 
\bm{\Y} = \Y^{a^1(m) \dots  a^l(m)} \, (\bvect_{a^1_1} \otimes \dots \otimes \bvect_{a^1_m})\otimes \dots \otimes(\bvect_{a^l_1} \otimes\dots \otimes \bvect_{a^l_m}),
\ee
where $a^j(m) \equiv \{ a^j_1 \dots a^j_m \}$ are groups of symmetric indices and the following anti-symmetry under $m$-wise exchange of indices holds,
\be 
\Y^{a^{\sigma(1)}(m) \dots  a^{\sigma(l)}(m)} = \sign(\sigma) \;\Y^{a^1(m) \dots  a^l(m)} \quad \forall \sigma \in S_{l}.
\ee
The symmetries of tensors such as $\X$ or $\Y$ are examples of so-called Plethysms (see e.g. \cite{Fulling:1992vm}).

More general plethysms can be constructed starting from a set of tensors $\bm{\T}_i$ with $i\in \{ 1,\dots ,k\}$ which each belong to the same symmetry class generated by some Young symmetrizer associated with some young tableau $\lambda$, $\Vspace_\lambda$. Considering now an other Young symmetrizer $\Youngsymantimu$ with $k$ boxes associated with a tableau $\mu$, we can construct the tensor $\bm{\T}_{\mu \circ \lambda}$ defined by
\be
\bm{\T}_{\mu \circ \lambda} = \Youngsymantimu \left( \bm{\T}_{1} \otimes \dots \otimes \bm{\T}_{k}\right),
\ee
where the action of the Young symmetrizer $\Youngsymantimu$ is just given as in the first line of (\ref{actiontensor}), where base vectors $\bvect_i$ are replaced by tensors $\bm{\T}_i$, i.e. it acts on places of tensors $\bm{\T}_i$ in the tensor product 
$\bm{\T}_{1} \otimes \dots \otimes \bm{\T}_{k}$. In components it can be defined by treating the collective indices of each tensor $\bm{\T}_{i}$ as a single unit. The tensor symmetry class corresponding to the symmetries of the tensor $\bm{\T}_{\mu \circ \lambda}$ is in general reducible (as a representation of the bisymmetric transformations - see appendices \ref{AppC} and \ref{AppD}). 

The symmetries 1. and 2. of section \ref{2.B} imply that $\bm{\E^m}$ has the symmetry of the plethysm 
\be
\plethysm, \label{Emplethysm}
\ee i.e. belong to the symmetry class $\Vspace_{ \plethysm}$, where the first plethysm entering in the tensor product, i.e. $\sym^m(\bigwedge^p)$ is associated with the symmetries 1. of section \ref{2.B} concerning indices with capital letters $A$ and $B_i$ (i.e. groups of p antisymmetric indices), and the second plethysm  $\sym^{(m-1)}(\sym^2)$  corresponds to symmetry 2. of section \ref{2.B} and is associated with lower case indices.

\section{Restrictions on the field equations from the symmetries 1. 2. and 3.} \label{section4}

After having shown how the different hypothesis made before lead to the conclusion that the tensor $\bm{\E^m}$ had the symmetries 1, 2 and 3, summarized at the end of section \ref{2.B}, we will further use these symmetries to characterize $\bm{\E^m}$ in a more detailed way. Our general method is just to decompose $\bm{\E^m}$ into pieces belonging to the irreducible components of the symmetry class defined by the symmetries 1. 2. and 3.  We first examine the consequence of $\bm{\E^m}$ having the symmetry 3. 

\subsection{Consequence of $\bm{\E^m}$ having the symmetry 3.}

In order to decompose $\bm{\E^m}$ into pieces belonging to the irreducible components of the symmetry class that it belongs to, one can act on $\bm{\E^m}$ with Young symmetrizers, as explained in appendix \ref{AppC}.  Consider such as symmetrizer built from a given Young diagram (see appendix \ref{AppB}) with a number of columns equal to or larger than three. Such a symmetrizer will be a linear combination of products of row group symmetrizers and column group antisymmetrizers (see equations (\ref{youngsymdef}-\ref{youngantidef})). If the number of column is larger or equal to three, such an operation (irrespectively on the choice made between the symmetric or antisymmetric presentation for the Young symmetrizers) will always involve a symmetrization on at least three space-time indices of the tensor $\bm{\E^m}$, which, as a consequence of the symmetry 3. of section \ref{2.B}, vanishes. Hence, we conclude that the only Young symmetrizers that can possibly enter into the decomposition of $\bm{\E^m}$ into irreducibles are those coming from diagrams with one or two columns at most. In the following, we will denote such a diagram appropriate to act on the tensor $\bm{\E^m}$, which has a variance $mp+2(m-1)$ (implying that the considered Young diagram should have $mp+2(m-1)$ boxes in total), as $(mp+2(m-1)-a,a)^t$ where 
$a$ is a positive integer parametrizing the length of the second column and $"t"$ denotes the transpose of the diagram with two lines, the first with $mp+2(m-1)-a$ boxes and the second with $a$ boxes (see appendix \ref{AppB}).

Having just shown that the irreducible representation of the symmetric group entering in the decomposition of the tensor symmetry class of $\bm{\E^m}$ are just those characterized by Young diagrams of the kind $(mp+2(m-1)-a,a)^t$, the next step is to determine the multiplicity $m_a$ of irreducibles each in one to one correspondence with a standard Young tableau built from the Young diagram $(mp+2(m-1)-a,a)^t$ entering into this decomposition. To do so we will use the machinery of Schur functions to find out this multiplicity inside the plethysm (\ref{Emplethysm}) as explained in appendix \ref{AppD}.


\subsection{Multiplicity of irreducibles with two columns Young diagrams inside $\plethysm$}
In order to determine the multiplicity $m_a$ as defined above, we will proceed in two steps. First, we will determine the multiplicity of irreducibles with two column diagrams inside the plethysms $\sym^m(\bigwedge^p)$ and $\sym^{(m-1)}(\sym^2)$ separately and then use the Littlewood Richardson rule to deal with the tensor product between these plethysms (see appendix \ref{LittlewoodRichardson}). This rule implies that indeed only diagrams with at most two columns inside each component $\sym^m(\bigwedge^p)$ and $\sym^{(m-1)}(\sym^2)$ can be combined to yield an irreducible corresponding to a Young diagram with no more than two columns inside the plethysm  $\plethysm$. We first consider the first factor of the tensor product $\sym^m(\bigwedge^p)$.

\subsubsection{Multiplicity of irreducibles with two columns Young diagrams inside $\sym^m(\bigwedge^p)$}
As explained in appendix \ref{AppD}, the Schur function corresponding to the symmetries of the plethysm $\sym^m(\bigwedge^p)$ is given by 
$ s_{(m)} \circ s_{(1)^p}(x)$ and can be decomposed as 
\be \label{schexpa}
s_{(m)} \circ s_{(1)^p} = \sum_\mu m_{\mu} s_\mu,
\ee
where $m_{\mu}$ gives the multiplicity of the irreducible representation characterized by the Young diagram $\mu$ inside the plethysm $\sym^m(\bigwedge^p)$. 
Here we are only interested in the multiplicities corresponding to the Young diagrams $\mu$ entering the above decomposition with at most two columns. This can be obtained as follows. First, we apply the $\Omega$ involution operation (see appendix \ref{AppD}) to the above decomposition \eqref{schexpa}. We get 
\be \label{schexpainv}
\Omega(s_{(m)}\circ s_{(1)^p}) = \sum_\mu m_{\mu} \Omega(s_\mu) = \sum_\mu m_\mu s_{\mu^t}.
\ee
Now we see that if we consider two particular variables, say $x_1$ and  $x_2$, among those upon which the Schur functions depend, the only monomials of the form $x_1^{\alpha_1} x_2^{\alpha_2}$(i.e. which only depend on $x_1$ and  $x_2$ and not on the others $x_i$, $i \neq 1,2$) which appear on the right hand side of the  above equation 
\eqref{schexpainv} all come from the Schur functions corresponding to Young diagrams $\mu^t$ with at most two rows (because any Schur function corresponding to a Young diagram with more than two rows necessarily depend on more than two variables). And these monomials and their coefficients will be the same if we just restrict ourselves to considering functions of just the two variables $x_1$ and  $x_2$ and set to zero all other contributions. This is what will be done in the following, enabling us 
to compute the multiplicities $m_\mu$ of the irreducibles corresponding to Young diagrams with at most two columns in the original decomposition \eqref{schexpa}.

Notice however, using  (\ref{OMEGAPARITY}), that  we have 
\be \label{schexpainv2}
\Omega(s_{(m)} \circ s_{(1)^p}) = \begin{cases}
s_{(m)}\circ \Omega(s_{(1)^p}) = s_{(m)} \circ s_{(p)} &\text{if $p$ is even}\\
\Omega(s_{(m)}) \circ \Omega(s_{(1)^p}) = s_{(1)^m} \circ s_{(p)}&\text{if $p$ is odd}
\end{cases} ,
\ee
hence, we shall distinguish the case of odd-p and even-p-forms.

\subsubsection{Case of an even-p-form: decomposition of $\left(s_{(m)} \circ s_{(p)}\right)(x_1,x_2)$}

First we write $s_{(p)}(x_1,x_2)$ explicitly, giving
\be \label{defsp}
s_{(p)}(x_1,x_2) = \sum_{r=0}^{p} x_1^{p-r} x_2^r ,
\ee 
and we define an ordered collection of variables $y_r$ by 
\be \label{defyr}
y_r = x_1^{p-r} x_2^r,\;\;{\rm for} \;\; 0\leq r \leq p.
\ee
 The \textit{plethysm} $s_{(m)} \circ s_{(p)}(x_1,x_2)$ is given by the Schur function $s_{(m)}(y_r(x))$, namely using (\ref{defschurm}),
\be \label{schexev}
s_{(m)} \circ s_{(p)}(x_1,x_2) = s_{(m)}(y_r(x)) = \sum_{0\leq r_1 \leq r_2 \leq \dots \leq r_m \leq p} y_{r_1} \dots y_{r_m}.
\ee
Note that the degree of $s_{(m)} \circ s_{(p)}(x_1,x_2)$ is $mp$. For a given $a$ verifying $0\leq a \leq mp$, the coefficient of the term $x_1^{mp-a}x_2^a$ in $s_{(m)}\circ s_{(p)}(x_1,x_2)$, which we denote as $C(x_1^{mp-a}x_2^a)$, can be deduced from \eqref{schexev} to be the number of (unordered) partitions of $a$ into $m$ non-negative integers within $\{0,1,2, \dots ,p  \}$ with repetitions allowed. Furthermore $s_{(m)} \circ s_{(p)}(x_1,x_2)$ has a unique expansion in terms of Schur functions that correspond to Young diagrams of at most two rows. Namely,
\be 
s_{(m)} \circ s_{(p)}(x_1,x_2) = \sum_{l=0}^{\left[{\frac{mp}{2}}\right]} m_{(mp-l,l)} s_{(mp-l,l)}(x_1,x_2),
\ee
where $[\dots]$ in the upper limit is the Gauss symbol, $(mp-l,l)$ denotes the Young diagram with with the first and second row of size $mp-l$ and $l$ respectively (in agreement with our notations of appendix \ref{AppB}), and $m_{(mp-l,l)}$ is the multiplicity of the Schur function 
\be
s_{(mp-l,l)}(x_1,x_2) = \sum_{b=l}^{mp-l}x_1^{mp-b}x_2^b,
\ee
that corresponds to this Young diagram. In order to determine these multiplicities, we just need to use the right hand side of the above equation to get 
\be 
C(x_1^{mp-a}x_2^a) = \sum_{l=0}^{{\rm min}\left(a,mp-a\right)} m_{(mp-l,l)}.
\ee
Now restricting to the case where $1\leq a \leq \left[{\frac{mp}{2}}\right]$ and using the above recursively we get that 
\be \label{mulmain1}
m_{(mp-a,a)} = C(x_1^{mp-a}x_2^a) - C(x_1^{mp-(a-1)}x_2^{a-1}),
\ee
while 
\bea
m_{(mp,0)} \equiv m_{(mp)} = C(x_1^{mp}) =1. 
\eea

\subsubsection{Case of an odd-p-form: decomposition of $s_{(1)^m}\circ s_{(p)}(x_1,x_2)$}
The decomposition can be performed by using the similar method as above. Starting as above from equations (\ref{defsp}) and (\ref{defyr}) and using (\ref{defschur1m})
we get
\be 
s_{(1)^m}\circ s_{(p)}(x_1,x_2) =s_{(1)^m}(y_r(x)) = \sum_{0 \leq r_1 < r_2 < \dots < r_m \leq p} y_{r_1} \dots y_{r_m}
\ee
 Notice that the crucial difference between this equation and \eqref{schexev} is that the variables $y_i$ have here to be distinct in the summation.
As before, for a given $a$ verifying $0\leq a \leq mp$, let us consider the coefficient $\bar{C}(x_1^{mp-a}x_2^a)$ of the term $x_1^{mp-a}x_2^a$ occurring in this expansion. We find this to be equal to the number of (unordered) partitions of $a$ into $m$ \textit{distinct} non-negative integers within $\{ 0,1,\dots, p\}$. Expressing now the plethysm $s_{(1)^m}\circ s_{(p)}(x_1,x_2)$ in the Schur function basis we have 
\be 
s_{(1)^m}\circ s_{(p)}(x_1,x_2) = \sum_{l=0}^{\left[ \frac{mp}{2}\right]} \bar{m}_{(mp-l,l)} s _{(mp-l,l)}(x_1,x_2),
\ee
where the notations are as before and we are after the multiplicities $\bar{m}_{(mp-l,l)}$. The coefficient $\bar{C}(x_1^{mp-a}x_2^a)$ verifies as above
\be 
\bar{C}(x_1^{mp-a}x_2^a) = \sum_{l=0}^{{\rm min}\left(a,mp-a\right)} \bar{m}_{(mp-l,l)}.
\ee
Now we can conclude as before that for  $1\leq a \leq \left[{\frac{mp}{2}}\right]$ 
\be \label{mulmain2}
\bar{m}_{(mp-a,a)} = \bar{C}(x_1^{mp-a}x_2^a) - \bar{C}(x_1^{mp-(a-1)}x_2^{a-1}),
\ee
while in this case we have
\bea
\bar{m}_{(mp,0)} \equiv \bar{m}_{(mp)} = \bar{C}(x_1^{mp}) =0. 
\eea

To summarize what has been achieved above, we have shown that the multiplicity $m_{(mp-a,a)^t}$ of the irreducible representation corresponding to the Young diagram $(mp-a,a)^t$ inside $\sym^m(\bigwedge^p)$ is given by the following formula 
\be \label{defNp}
m_{(mp-a,a)^t} = \begin{cases}
N^p_{a,m} - N^p_{a-1,m}&\text{if $p$ is even}\\
N^{p,distinct}_{a,m} - N^{p,distinct}_{a-1,m}&\text{ if $p$ is odd}
\end{cases},
\ee
where for $r \geq 0$, $N^p_{r,s}$ is the number of (unordered) partitions of $r$ into $s$ non-negative integers within $\{0,\cdots,p\}$ with repetitions allowed and $N^{p,distinct}_{r,s}$ is the number of (unordered) partitions of $r$ into $s$ distinct non-negative integers within $\{0,\cdots,p\}$; and we have 
$N^p_{r,s} =0 $ and $N^{p,distinct}_{r,s} =0$ for negative $r$.


\subsubsection{Multiplicity of irreducibles with two columns Young diagrams inside $\sym^{(m-1)}(\sym^2)$}
We proceed as above first applying the $\Omega$ involution operation to the Plethysm of Schur functions $s_{(m-1)} \circ s_{(2)}$. 
We have 
\be 
\Omega \left( s_{(m-1)} \circ s_{(2)}\right) = s_{(m-1)} \circ s_{(1)^2}.
\ee
Then we restrict the variables to $(x_1,x_2)$ and get for the second factor above 
\be 
s_{(1)^2}(x_1,x_2) = x_1 x_2.
\ee
Thus we have only one variable $y_i$ to consider in the following step. Hence we get the very simple decomposition
\be 
s_{(m-1)} \circ s_{(1)^2}(x_1,x_2) = x_1^{m-1}x_2^{m-1} = s_{(m-1,m-1)}(x_1,x_2),
\ee
which leads us to conclude that inside $\sym^{(m-1)}(\sym^2)$ there is only one irreducible corresponding to Young diagrams with up to two columns and it is characterized by the Young diagram with two columns of equal size equal to $(m-1)$, i.e. the diagram $(m-1,m-1)^t$.
\label{subsub4}

\subsubsection{Final result: application of the Littlewood Richardson rule}
\label{subsub5}
In order to obtain the multiplicity we are after we just need to consider the tensor product of a given two column diagram  $(mp-a,a)^t$ corresponding to one given irreducible inside $\sym^m(\bigwedge^p)$ by the two column diagram $(m-1,m-1)^t$ (which itself corresponds to the only irreducible with at most two columns inside $\sym^{(m-1)}(\sym^2)$ indexed by the diagram 
$(m-1,m-1)^t$), and focus only on the Young diagram which have one or two columns. A very simple application of the Littlewood Richardson rule yields only one such diagram: the one where the first column of $(m-1,m-1)^t$ is glued at the end of the first column of the diagram $(mp-a,a)^t$ and where the second column is glued at the end of the second column of 
$(mp-a,a)^t$ (this is because the Littlewood Richardson rule imposes that no boxes stemming from the same row can be glued to a column where a similar box has already been glued).
Hence for one given irreducible indexed by the diagram  $(mp-a,a)^t$ inside $\sym^m(\bigwedge^p)$ we get only one irreducible inside the plethysm $\plethysm$ indexed by a diagram 
$(m(p+1)-a-1,a+m-1)^t$. Hence, by changing $a+m-1$ into $a$, we get the multiplicity $M_{(mp + 2(m-1)-a,a)^t}$ of the irreducible corresponding to the Young diagram $(mp + 2(m-1)-a,a)^t$ inside the plethysm $\plethysm$ given by 
\be \label{finalrescount}
M_{(mp + 2(m-1)-a,a)^t} = m_{(mp-a+m-1,a-m+1)^t} =
\begin{cases}
N^p_{a-m+1,m} - N^p_{a-m,m}&\text{if $p$ is Even}\\
N^{p,distinct}_{a-m+1,m} - N^{p,distinct}_{a-m,m}&\text{ if $p$ is Odd}
\end{cases},
\ee
where the numbers $N^p_{r,s}$ and $N^{p,distinct}_{r,s}$ are defined as in equation (\ref{defNp}). Note in particular that in order for $N^{p,distinct}_{r,s}$ to be non zero we need obviously to have $s \leq p+1$ (otherwise there are less integers inside $\{0,\cdots,p\}$ than necessary to get a partition of $r$ into $s$ distinct such integers), this translate for the odd $p$ case to the very important condition 
\be \label{Condmp}
m \leq p+1.
\ee
The origin of this condition can be traced back to the step {\it 3} of the above derivation: when condition (\ref{Condmp}) is not fulfilled there are less variables $y_r$ than the number of boxes in the single column of the Young diagram $(1)^m$ and hence one cannot build the Schur function $s_{(1)^m}(y_r(x))$.

\section{Explicit construction of p-form Galileon theories} \label{section5}

The above derived restrictions on the tensor $\bm{\E^m}$ allows to count the maximum number of possible theories of Galileon p-forms with specific properties. We will differ the full analysis of this question to an other work but we would like here to consider two particular cases, the one of a vector (corresponding to $p=1$) and the first non trivial case of an odd-p-form with $p=3$.

\subsection{Vector Galileon: no-go}
The case $p=1$ has already been studied in \cite{no-go} where it was shown that no such non trivial Galileon theory existed. This results can be obtained here in a very simple way using the above results. Indeed, for $p=1$ we find from (\ref{Condmp}) that we must have $m \leq 2$ in order to have a non vanishing multiplicity $(mp + 2(m-1)-a,a)^t$ inside the plethysm $\plethysm$. Recalling that  $(m-1)$  gives the order of the taken derivatives of the field equation $\bm{\E}$ with respect to the second derivative of the gauge field, we find that, under our hypotheses, the field equations can only depend on the second derivative of the gauge field linearly which matches the results of \cite{no-go}.

\subsection{A non trivial 3-form theory}

The next interesting case for odd $p$ corresponds to $p=3$ which we investigate here. In this case, obviously from (\ref{Condmp}), we must have $m \leq 4$ in order to have a non vanishing interesting tensor $\bm{\E^m}$. Moreover, when $p=3$ the rank of this tensor is 
\bea \label{ranktensor}
(mp+2(m-1)) = 5 m -2.
\eea
  By taking large enough $m$ (i.e. enough number of derivatives with respect of the second derivatives of the $p-$form), and since we are looking at theories where the field equations only depend on second derivatives of the form, we should be able to end up with a tensor $\bm{\E^{m_{max}}}$ which is solely built from the metric $\eta^{ab}$ and the $\bm{\epsilon}$ tensor which are the only tensors available besides the second derivatives of the form. Here we focus on the first case, and one must then have $\bm{\E^{m_{max}+1}}=0$.
However, $\eta^{ab}$ has obviously an even rank, this should hence be true as well for the tensor  $\bm{\E^{m_{max}}}$ obtained as we just explained. Hence, in the $p=3$ case, we are left with the only two possibilities $m_{max} = 2$ and $m_{max}= 4$ (see equation (\ref{ranktensor}) and (\ref{Condmp})). The first possibility does not lead to any interesting Galileon-like theory of a 3-form since in that case $\bm{\E^{3}}$ vanishes which means again that the field equations are at most linear in second derivatives of the 3-form. So the only left over case is for $m_{max}=4$. In this case, we have from (\ref{ranktensor}) that $\bm{\E^{4}}$ has rank 18 and should only depend on $\eta^{ab}$.

The starting point of our construction is hence the tensor product $\bm{\eta_{9}} = \eta \otimes \dots \otimes \eta$ (9 factors), from which we want to build a tensor having all the required symmetries 1. 2. and 3. summarized in section \ref{2.B}. The tensor $\bm{\eta_{9}}$ belong by construction to the symmetry class of the plethysm $\sym^9(\sym^2)$ and using the results of section \ref{subsub4} we see that there is only one irreducible inside this plethysm corresponding to a two column diagram with two columns of equal size. 
It is easy to see how the corresponding irreducible representation of $S_{18}$ can be constructed. Indeed, we can consider the Young symmetrizer $\Younganti$ where $\lambda_k$ corresponds to the standard filling of the two column diagram $(2)^9$ where a given row contains the pair of indices of one given factor $\eta$ of the tensor product $\bm{\eta_{9}}$. 
The action of $\Younganti$ on  $\bm{\eta_{9}}$ then simply gives the tensor $\bm{\epsilon_{2(18)}}= \bm{\epsilon_2}$ (where here and below, we consider the simplest case where space-time has the required minimum D=9 number of dimensions, see eq. (\ref{DEFAten})) up to an overall constant, i.e. we have $\Younganti \bm{\eta_{9}} = -2^9\cdot 9!\cdot\bm{\epsilon_2}$, which can be seen using the identity 
\ba
\sum_{\sigma \in S_9} {\rm sign}(\sigma)
\eta^{a_{\sigma(1)}b_{\vphantom{()}1}}
\eta^{a_{\sigma(2)}b_{\vphantom{()}2}} \ldots
\eta^{a_{\sigma(9)}b_{\vphantom{()}9}} &=&
- \epsilon^{a[9]}\,
\epsilon^{b[9]} 
= - \epsilon_2^{a[9]b[9]}.
\ea

Now, because $\bm{\epsilon_2}$ is obviously non vanishing, we can use the statements of appendix \ref{AppC1} showing that 
$\mathbb{R}[S_{18}]\bm{\epsilon_2} \equiv \bm{M_{\epsilon_2}}$ provides an irreducible representation of $S_{18}$ indexed by the Young diagram $(2)^9$.
Using the results of section \ref{subsub5}  we see that the plethysm of interest here, given by (for $m=4$ and $p=3$)
$\plethysmpart$, contains one and only one equivalent representation indexed by the same diagram $(9,9)^t=(2)^9$. This follows from equation (\ref{finalrescount}) with $a=9, m=4, p=3$ yielding $N^{3,distinct}_{6,4}=1$ and $N^{3,distinct}_{5,4}=0$ simply because $6=0+1+2+3$.
Let us consider one given non vanishing tensor $\T$ belonging to this irreducible and to the symmetry class $\plethysmpart$, then, using results summarized in appendices \ref{AppC1} and \ref{AppC2} we have that this irreducible is just given by $\bm{M_{\T}}$.
Schur's lemma and its corollaries exposed in appendix \ref{AppA3} then show that there is a unique (up to scalar multiplication) $S_{18}$-isomorphism between  $\bm{M_{\T}}$ and $\bm{M_{\epsilon_2}}$, which we call $\psi$. Our next aim is to use $\psi$ to build the explicit form of a suitable non vanishing $\bm{\E^{4}}$ inside $\plethysmpart$.

We first note that if we consider a given $S_{18}$-isomorphism $\phi$ between  $\bm{M_{\T}}$ and $\bm{M_{\epsilon_2}}$, this map is entirely defined by the image of the tensor  $\bm{\T}$, $\phi\left(\bm{\T}\right)$. Indeed, once this image is defined, the image of any tensor belonging to $\bm{M_{\T}}$ defined by $\sum_{\sigma\in S_{18}} f(\sigma) \sigma \left( \bm{\T}\right)$, is just given by (using the notations of appendix \ref{AppB1} as well as the fact that $\phi$ is an intertwiner)
\ba
\phi\left(\sum_{\sigma\in S_{18}} f(\sigma) \sigma \left( \bm{\T}\right)\right) = \sum_{\sigma\in S_{18}} f(\sigma) \sigma \left(\phi\left( \bm{\T}\right)\right).
\ea 
Hence, we see that a complete set of intertwiners for the vector space of intertwiners between $\bm{M_{\T}}$ and $\bm{M_{\epsilon_2}}$ is just given by the intertwiners $\phi_\sigma$ defined by 
\ba \label{defphisigma}
\phi_\sigma\left(\bm{\T} \right) = \sigma \left(\bm{\epsilon_2}\right), \quad
\sigma \in S_{18}.
\ea
Namely, any intertwiner between $\bm{M_{\T}}$ and $\bm{M_{\epsilon_2}}$ can be written as $\sum_{\sigma\in S_{18}} F(\sigma) \phi_{\sigma}$, where $F(\sigma)$ is a function on $S_{18}$.

Let us then choose a specific order and labeling for the $18$ indices of $\bm{\T}$ which corresponds to the order of the indices of $\bm{\E^4}$ given in equation (\ref{defEm}) for $m=4$ and $p=3$. As $\bm{\T}$ belongs to $\plethysmpart$, it has in particular the symmetries 1. and 2. defined in section \ref{2.B}. In order to picture these symmetries here it turns out easier to first define (for generic $p$ and $m$) operators ${\cal O}_1^{(m,p)}$ and ${\cal O}_2^{(m)}$ acting as defined below on an arbitrary tensor $\bm{\X}$.
In order to simplify notations, we first define the following shortcut $\X_{comp}^{(m,p)}$ to denote the component of the tensors $\bm{\X}$  with the following specific labeling
\ba
\X_{comp}^{(m,p)} \equiv \X^{a\{p\} b_1\{p\}  c_1d_1 \dots b_{m-1}\{p\} c_{m-1}d_{m-1}}, 
\ea
where $a\{p\}=\{a_1,\cdots,a_p\}$ and $b_i\{p\}=\{b_{i,1},\cdots,b_{i,p}\}$. The operators ${\cal O}_1^{(m,p)}$ and ${\cal O}_2^{(m)}$ are then defined as 
\ba \label{defoper} 
\left({\cal O}_1^{(m,p)} \left(\bm{\X}\right)\right)^{A B_1 c_1d_1 \dots B_{m-1} c_{m-1}d_{m-1}}  &\equiv& \Symop_{(a\{p\}, b_1\{p\}, \dots, b_{m-1}\{p\})} \Antisymop_{(a\{p\})} \prod_{i=1}^{m-1}\Antisymop_{(b_i\{p\})} \left(\X_{comp}^{(m,p)}\right),
\\
\left({\cal O}_2^{(m)} \left(\bm{\X}\right)\right)^{a\{p\} b_1\{p\} c_1d_1 \dots b_{m-1}\{p\} c_{m-1}d_{m-1}}    &\equiv&  \Symop_{\left(\{c_1,d_1\},\dots , \{c_{m-1},d_{m-1}\}\right)} \prod_{i=1}^{m-1}\Symop_{(c_i,d_i)} \left(\X_{comp}^{(m,p)}\right), 
\ea
where $A=a[p]$ and $B_i=b_i[p]$ are sets of antisymmetric indices made respectively of $a\{p\}$ and $b_i\{p\}$, $\Symop_{(\alpha_1, \dots, \alpha_k)}$ means a symmetrization on the arguments $\alpha_1, \dots, \alpha_k$ and $\Antisymop_{(\alpha_1, \dots, \alpha_k)}$ an antisymmetrization on the arguments $\alpha_i$, these arguments being either space-time indices (e.g. when $\{\alpha_1, \dots, \alpha_p\} = a\{p\} = \{a_1,\dots, a_p\}$) or list of indices (e.g. when $\{\alpha_1, \dots, \alpha_{m}\} = \{a\{p\},b_1\{p\},\dots, b_{m-1}\{p\}\}$). 
One then has
\ba
{\cal O}_1^{(4,3)} \left(\bm{\T}\right) &=& \bm{\T},  \\
{\cal O}_2^{(4)} \left(\bm{\T}\right) &=& \bm{\T}, 
\ea
as a consequence of the fact that $\bm{\T}$ has the symmetries 1 and 2. 
Note that these operators commute (as they do not act on the same indices) and are also idempotent elements (once properly normalized) of the group algebra $\mathbb{R}[S_{18}]$. 
Their product acts as a projector on the plethysm $\plethysm$ and constitutes a generating idempotent of the tensor symmetry class corresponding to this plethysm as defined in (\ref{genidemvect}). Let us then introduce $\sigma_0\in S_{18}$ so that it maps $\bm{\epsilon_2}$ to 
\ba
\left(\sigma_0\left(\bm{\epsilon_2}\right)\right)^{a\{3\} b_1\{3\}  c_1d_1 b_2\{3\} c_2d_2  b_3\{3\} c_{3}d_{3}} = \epsilon_2 ^{c\{3\} a\{3\} b_{1,1}b_{1,2}b_{2,1} d\{3\} b_{1,3}b_{2,2}b_{2,3}b_3\{3\}},
\ea
which corresponds to the filling $\lambda_0$ of the Young tableau $(2)^9$. Here, $\lambda_0$ is defined by
\begin{center}
\ytableausetup 
{mathmode, boxsize=0.7cm, centertableaux}
$\lambda_0^t$ = \begin{ytableau}
c_1 & c_2 & c_3 & a_1 & a_2 & a_3 & b_{1,1} &
b_{1,2} &
b_{2,1} \\
d_1 & d_2 & d_3 &  b_{1,3} &  b_{2,2}& b_{2,3}& b_{3,1}& b_{3,2}& b_{3,3}
\end{ytableau},
\end{center}
and we have defined $a\{3\} = \{a_{1},a_{2},a_{3}\}$, $c\{3\} = \{c_{1},c_{2},c_{3}\}$, $d\{3\} = \{d_{1},d_{2},d_{3}\}$ and $b_i\{3\} = \{b_{i,1},b_{i,2},b_{i,3}\}$ for $1 \leq i \leq 3$. Concretely, $\sigma_0\in S_{18}$ is defined as $\sigma_0(i)=5i+2$, $\sigma_0(3+i)=i$, $\sigma_0(7)=4$, $\sigma_0(8)=5$, $\sigma_0(9)=9$, $\sigma_0(9+i)=5i+3$, $\sigma_0(13)=6$, $\sigma_0(14)=10$, $\sigma_0(15)=11$, $\sigma_0(15+i)=13+i$, where $i=1,2,3$.

Let us then consider the intertwiners $\phi_{\sigma}$ defined above as in (\ref{defphisigma}).
Because $\phi_{\sigma}$ is an intertwiner, one has 
\ba \label{actionO1O2}
{\cal O}_1^{(4,3)} {\cal O}_2^{(4)} \left( \phi_\sigma\left(\bm{\T} \right) \right) = \phi_\sigma\left({\cal O}_1^{(4,3)} {\cal O}_2^{(3)}\left(
\bm{\T} \right)\right)= \phi_\sigma\left(
\bm{\T} \right) = {\cal O}_1^{(4,3)} {\cal O}_2^{(4)} \left(\sigma \left(\bm{\epsilon_2}\right)\right)  
\ea
where the first equality follows from the fact $\phi_{\sigma}$ is an intertwiner and the second from the fact $\T$ belongs to the symmetry class of $\plethysm$.
Now it is easy to see that for any $\sigma$ such that $\sigma^{-1} \sigma_0$ belong to the column group of the diagram $\lambda_0$ (i.e is such that the filling of the tableau $(2)^9$
corresponding to $\sigma \left(\bm{\epsilon_2}\right)$  and the filling corresponding to $\sigma_0 \left(\bm{\epsilon_2}\right)$ only differ by the order of the indices in each columns) is such that $\sigma \left(\bm{\epsilon_2}\right)$ and $\sigma_0 \left(\bm{\epsilon_2}\right)$ are equal up to a factor $\pm 1$ due to the antisymmetry of the two factor $\bm{\epsilon}$ entering into the definition of $\bm{\epsilon_2}$. On the other hand, whenever $\sigma^{-1} \sigma_0$ does not belong to the column group of the diagram $\lambda_0$, one has ${\cal O}_1^{(4,3)} {\cal O}_2^{(4)} \left(\sigma \left(\bm{\epsilon_2}\right)\right) =0$. Indeed, in this case, the filling of the tableau $(2)^9$
corresponding to $\sigma \left(\bm{\epsilon_2}\right)$ is such that either (i) one column contains strictly more than three indices in the set $c\{3\} \cup d\{3\}$ (and hence at least two of these indices $c_i$ and $d_i$ carry the same subindex $i$ in which case $\sigma \left(\bm{\epsilon_2}\right)$ is annihilated by the action of ${\cal O}_2^{(4)}$)
or if it not the case one can first assume that, due to the action of the symmetrization inside ${\cal O}_2^{(4)}$ all index in $c\{3\}$ and all index in $d\{3\}$ of $\sigma \left(\bm{\epsilon_2}\right)$ are in a different column as in $\lambda_0$ and then either (ii) at least one column contains the same number of indices each belonging to two sets among $\left\{a\{3\},b_1\{3\},b_2\{3\},b_3\{3\}\right\}$  (in which case $\sigma \left(\bm{\epsilon_2}\right)$ is annihilated by the combined action of ${\cal O}_2^{(4)}$ and  ${\cal O}_1^{(4,3)}$) or (iii) this is not the case in which case the action of  ${\cal O}_2^{(4)} {\cal O}_1^{(4,3)}$ on $\sigma \left(\bm{\epsilon_2}\right)$ is just  proportional to 
${\cal O}_2^{(4)} {\cal O}_1^{(4,3)} \left(\sigma_0 \left(\bm{\epsilon_2}\right)\right)$.

As a consequence of the above, of equation (\ref{actionO1O2}), and of the fact we know that there exists a unique (up to a constant) non vanishing intertwiner between $\bm{M_{\T}}$ and $\bm{M_{\epsilon_2}}$, we reach the important conclusion that ${\cal O}_1^{(4,3)} {\cal O}_2^{(4)} \left(\sigma_0 \left(\bm{\epsilon_2}\right)\right)$
cannot vanish. Hence we can set $\bm{\E^4}$ to be equal to 
\ba
\left(\bm{\E^4}\right)^{A B_1  c_1d_1 B_2 c_2d_2   B_3 c_{3}d_{3}} = {\cal O}_1^{(4,3)} {\cal O}_2^{(4)} \left(\epsilon_2 ^{c\{3\} a\{3\} b_{1,1}b_{1,2}b_{2,1} d\{3\} b_{1,3}b_{2,2}b_{2,3} b_3\{3\}} \right),
\ea
where $A=a[3]$ and $B_i=b_i[3]$ are sets of antisymmetric indices made of $a\{3\}$ and $b_i\{3\}$, respectively. It is easy to see explicitly that this tensor does not vanish. For this purpose it suffices to show that the component with $c_i=d_i=i$, $a_i=b_{3,i}=3+i$, $b_{1,i}=b_{2,i}=6+i$ ($i=1,2,3$) is non-vanishing. First, the operator ${\cal O}_2^{(4)}$ acted on this component is equivalent to the multiplication by $3!\cdot 2^3$. Second, $\Antisymop_{(a\{p\})}\Antisymop_{(b_3\{p\})}$ is equivalent to the multiplication by $(3!)^2$. Third, in order to show that $\Symop_{(a\{p\}, b_1\{p\}, b_2\{p\}, b_3\{p\})}$ is also reduced to a multiplication by a non-vanishing constant, let us rewrite it as a simple sum over $\alpha\in S_4$ acted on labels $\{1,2,3,4\}$, where the label $1$ corresponds to $a\{p\}$ and the labels $1+i$ correspond to $b_i\{p\}$ ($i=1,2,3$). Then, for the specific component that we are currently considering it is obvious that $\Symop_{(a\{p\}, b_1\{p\}, b_2\{p\}, b_3\{p\})}$ is reduced to the summation over $\{\alpha\in S_4 \mbox{ s.t. } (\alpha(1), \alpha(4)) = (1,4)\mbox{ or }(4,1)\mbox{ or }(2,3)\mbox{ or }(3,2) \}$, and thus is equivalent to the multiplication by $4\cdot 2$. Finally, in order to show that $\Antisymop_{(b_1\{p\})}\Antisymop_{(b_2\{p\})}$ is also a multiplication by a non-vanishing constant, let us rewrite it as a weighted sum over $\beta\in S_3$ and $\gamma\in S_3$ with the weight $\sign(\beta)\sign(\gamma)$. For the component of interest it is then easy to see that $\Antisymop_{(b_1\{p\})}\Antisymop_{(b_2\{p\})}$ is reduced to the simple sum over $\beta=\gamma\in S_3$ followed by the multiplication by $2$, where the factor $2$ takes care of the interchange between $\gamma(1)$ and $\gamma(2)$ relative to $\beta(1)$ and $\beta(2)$. Thus $\Antisymop_{(b_1\{p\})}\Antisymop_{(b_2\{p\})}$ is equivalent to the multiplication by $2\cdot 3!$. This completes the explicit proof that the tensor $\bm{\E^4}$ defined above does not vanish.

To get the equation of motion we integrate this tensor three times with respect to the second derivative of the 3-form and thus we get (omitting integration constants which will be discussed in a future publication as well as symmetrizations antisymmetrizations inside the operators ${\cal O}_1^{(4,3)} {\cal O}_2^{(4)}$ which are redundant with the contraction with the second derivatives of the form) 
\be \label{field3}
\E^A = \left(\Symop_{\left(A,b_1\{3\},b_2\{3\},B_3\right)} \epsilon^{c\{3\} A b_{1,1} b_{1,2} b_{2,1}} \epsilon^{d\{3\} b_{1,3} b_{2,2} b_{2,3} B_3}\right) \A_{B_1,c_1 d_1}\A_{B_2,c_2 d_2}A_{B_3,c_3 d_3}.
\ee
Here, as already stated at the beginning of section~\ref{2.B}, we have not introduced dependence of the integration ``constants'' on first derivatives of the p-form for simplicity. Also as already stated, the integration ``constants'' cannot depend on the p-form itself without derivatives acted on it because of (\ref{G1}). These equations of motion derive from the following action (up to an overall constant and a total derivative)
\be 
S = \int_{\cal M} \epsilon^{c\{3\} A b_{1,1} b_{1,2} b_{2,1}} \epsilon^{d\{3\} b_{1,3} b_{2,2} b_{2,3} B_3} \A_A \A_{B_1,c_1 d_1}\A_{B_2,c_2 d_2}A_{B_3,c_3 d_3}
\ee
where the left over operator $\Symop_{\left(A,b_1\{3\},b_2\{3\},B_3\right)}$ has just been removed as keeping it would yield an equivalent action up to a boundary term. 
Hence we have obtained a non trivial Galileon like theory for a 3 form necessitating at least $9$ space-time dimensions. This theory fulfill our initial criteria: it has an action principle, has field equations (\ref{field3}) which only contain second derivatives and which are gauge invariant. The gauge invariance, although guaranteed by our formalism,  can also be explicitly checked from the field equations (\ref{field3}). Indeed, any replacement there of one $\bm{\A}$ by the exterior derivative $d \bm{\B}$ of a two form $\bm{\B}$ yield a vanishing expression because the index of the derivative coming from the operator $d$ is contracted with one epsilon tensor $\bm{\epsilon}$ which also contains one of the index of the second derivatives acting on the replaced $\bm{\A}$ in equation (\ref{field3}). It is interesting also to stress that the action is only gauge invariant up to a boundary term which in fact makes our theory similar to Chern-Simon's. 
To further investigate this question, one notices that the above action can be written after suitable integration by part and relabeling as (leaving aside an overall sign of combinatorial origin)
\be 
S = \int_{\cal M} d^9 x \epsilon^{a[9]} \epsilon^{b[9]} \left(\partial_{a_1}\F_{B_1} \right)\left(\partial_{b_1} \F_{A_1}\right) \left( \partial_{a_6} \A_{b_7 b_8 a_9} \right) \left(\partial_{b_6} \A_{a_7 a_8 b_9}  \right)
\ee
where here $B_1= \{b_2,b_3,b_4,b_5\}$ and $A_1 =\{a_2,a_3,a_4,a_5\}$, $\A$ is by assumption a 3 form and $\F = d \A$ is the associated field strength. 
This can be contrasted with the action found in \cite{Deffayet:2010zh} which was reminded in the introduction, Eq. (\ref{gefdef3}) where the structure of the index contraction is different for what concerns the first derivatives.

\section{Conclusion} \label{section6}

In this work we have investigated Galileon like p-form theories and provided a first step towards their full classification. 
Focusing on the case of single p-forms with gauge invariant pure second order field equations deriving from an action principle, we have exposed a method to get an upper bound on the number of such non trivial theories. This allowed us in particular to give a new proof of the no-go theorem obtained in \cite{no-go} for gauge invariant vector Galileons corresponding to p$=1$. We also constructed explicitly a non trivial theory for a single 3-form, making explicit that this no-go theorem does not extend to higher odd p. This work can obviously be extended in various directions. First, it should be possible to compute the multiplicities obtained at the end of section \ref{section4} for all single p-form theories and say "reasonable" space-time dimensions (say e.g. $D\leq 11$) and possibility to construct all of them following the method exposed here. In particular one should be able to investigate the issue of uniqueness of the even-p-form theories found in \cite{Deffayet:2010zh}. A second direction of investigation is to covariantize these theories in the spirit of what has been done in \cite{Deffayet:2009wt} for the scalar case and in \cite{Deffayet:2010zh} for p-forms.

\section*{Acknowledgments}
We thank Gilles Esposito-Far\`ese for interesting and stimulating discussions.
The work of C.D.~and V.S.~was supported by the European Research Council
under the European Community's Seventh Framework Programme
(FP7/2007-2013 Grant Agreement no.~307934, NIRG project).
The work of S.M. was supported in part by Grant-in-Aid for Scientific Research 24540256 and the WPI Initiative, MEXT Japan. Part of his work has been done within the Labex ILP (reference ANR-10-LABX-63) part of the Idex SUPER, and received financial state aid managed by the Agence Nationale de la Recherche, as part of the programme
Investissements d'avenir under the reference ANR-11-IDEX-0004-02. He is thankful to colleagues at Institut d'Astrophysique de Paris for warm hospitality. 

\appendix 
\section{Some elements about real and complex linear representations of finite groups}

Let $\Vspace$ be a vector space over the field of real numbers $\mathbb{R}$ or the field of complex numbers $\mathbb{C}$. We note as usual  $\mathbf{GL}(\Vspace)$ the vector space of linear and invertible endomorphisms of $\Vspace$ (i.e. of automorphisms of $\Vspace$), and we spell out in this appendix some standard (and some less standard) results and definitions 
 about linear representation theory of finite group that we will use in this work (see e.g. \cite{Serre}).
\subsection{Generalities} \label{RepGen}
Suppose $G$ is a finite group. A \textit{\textbf{linear representation}} of $G$ into $\Vspace$  is a homomorphism, $\rho: G \to \mathbf{GL}(\Vspace)$. I.e. $\rho$ is a map from  $G$ to $\mathbf{GL}(\Vspace)$ which preserves the group structure and hence verifies 
\ba 
\rho(g_1g_2) = \rho(g_1) \rho(g_2) \quad {\rm for \;\; all}\;\; g_1, g_2 \in \mathbf{G}
\ea
Below, we will sometimes also use the notation $\rho_g$ to denote the image of $g$ by $\rho$, $\rho(g)\equiv \rho_g$ (such that one has $\rho_{g_1g_2} = \rho_{g_1} \rho_{g_2}$) as well as the more correct "$\left(\Vspace,\rho\right)$"  to designate the representation of $G$ under consideration.

Consider $(\Vspace, \rho), (\Wspace, \rho')$, two representations of the group $G$. A linear map $\phi:\Vspace \to \Wspace$,  that commutes with the group action, i.e. such that  
\be 
\rho'_g \phi(\bm{\mathcal V}) =  \phi(\rho_g \bm{\mathcal V}) \quad \forall g \in G \;\; {\rm and} \;\; \forall \bm{\mathcal V} \in \bm{V}
\ee
is called an \textit{\textbf{intertwiner}} map from $\Vspace$ to  $\Wspace$ (such a map is also said to be  \textit{\textbf{equivariant}} or also a  \textit{\textbf{G-map}} or a  \textit{\textbf{G-homomorphism}}).

Let $\Vspace,\Wspace$ be vector spaces of equal dimension and $\phi:\Vspace \to \Wspace $ be a vector space isomorphism. Given a representation $\rho$ of group $G$ on $\Vspace$, the map $\phi$ induces a unique representation $\rho'$ of $G$ on $\Wspace$ via 
\be 
\rho_g' = \phi \circ \rho_g \circ \phi^{-1} \quad g \in G
\ee
The two representations $\rho$ and $\rho'$ are said to be \textit{\textbf{equivalent}}.

Let $\Uspace\subset \Vspace$ be a vector subspace of $\Vspace$ and $\rho$ be a representation of $G$ acting on $\Vspace$. We say that $\Uspace$ is an \textit{\textbf{invariant subspace}} of $\Vspace$ under $\rho$ (or that $\Uspace$ is stable under $\rho$), iff
\be 
\rho(g) \bm{\mathcal V} \in \Uspace \quad \forall \bm{\mathcal V} \in \Uspace, \forall g \in G
\ee 
Obviously, then $\rho$ provides a representation of G into the vector space $\bm{U}$.
If $\Uspace $ is such that it does not contain any subvector space other than itself (and the trivial $\{0\}$) that is stable under $G$, we say, $\Uspace$ furnishes an \textit{\textbf{irreducible representation}} of $G$.

An important theorem states that \textit{\textbf{any representation of a finite group is a direct sum of irreducible representations}}.
I.e. the vector space $\Vspace$ defining the representation under consideration can always be written as 
\be \label{decomp}
\bm{V} = \bigoplus_i \bm{W}_i
\ee 
where, $\bm{W}_i$ are irreducible representations of the group $G$. Another way to state this result is that any representation of a finite group is \textit{\textbf{completely reducible.}}
Note that this decomposition is not unique, however any other similar decomposition gives a decomposition into irreducible representations.
Moreover, some of the $\bm{W}_i$ entering the direct sum in (\ref{decomp}) can give equivalent representations. Hence, we can denote by $\bm{\bar{{W}}}_j$ the vector spaces entering into the direct sum (\ref{decomp}) which furnish inequivalent representations (i.e. such that $\bm{\bar{W}}_i$ and  $\bm{\bar{W}}_j$ are \textit{inequivalent} irreducible representations whenever $i \neq j$). This implies that for a given irreducible representation $\bm{\bar{{W}}}_j$, there are $m_j$ (with $m_j \geq 1$) equivalent irreducibles entering into the direct sum (\ref{decomp}), which we can write as 
\be 
\Vspace = \bigoplus_j \bm{\bar W}^{\oplus m_j}_j.
\ee
The number  $m_j$ of vector spaces $\bm{\bar{W}}_j$ contained in $\Vspace$ corresponding to the same irreducible up to equivalence is called the \textbf{\textit{multiplicity}} of $\bm{\bar{W}}_j$.

A simple lemma given below will also be used in the following: let us consider two representations $(\Vspace,\rho), (\Wspace,\rho')$ and a G-homomorphism $\psi$ between them, and assume that $\Vspace$ is irreducible. Then (i) the image $Im(\Vspace)$ of $\Vspace$ by $\psi$ is an invariant subspace of $\Wspace$ and (ii) if $Im(\Vspace)$ does not reduce to ${0}$ it is irreducible. Indeed, consider first an arbitrary element $\bm{\mathcal W}$ of $Im(\Vspace)$, then there exists $\bm{\mathcal V}\in\Vspace$ such that $\bm{\mathcal W} = \psi (\bm{\mathcal V})$.  For an arbitrary element $g$ of $G$, one thus has $\rho'_g\left(\bm{\mathcal W}\right) = \rho'_g\left(\psi (\bm{\mathcal V})\right)=\psi(\rho_g(\bm{\mathcal V}))$. But since $\Vspace$ is irreducible, then it is also invariant and hence $\rho_g(\bm{\mathcal V}) \in \Vspace$, which proves (i). Similarly, consider now an invariant subspace $\Wspace'$ inside $Im(\Vspace)$, then it is easy to see its reciprocal image  $\psi^{-1}\left(\Wspace'\right)$  is an invariant subspace of $\Vspace$. However, since we have assumed that $\Vspace$ is irreducible, its only non trivial invariant subspace is itself, which means that one must have $\Wspace' = Im(\Vspace)$ which ends the proof of (ii). 

\label{AppA1}

\subsection{Absolute Irreducibility} \label{AppA2}
Let us now assume further that the base field of $\Vspace$ is the field of real numbers $\mathbb{R}$ (and, we stress, not its algebraic closure $\mathbb{C}$). In order to make this clear we denote as $\VspaceR$ the vector space $\Vspace$ with  $\mathbb{R}$ as a base field. We assume that $(\VspaceR,\rho)$ is a representation of the group $G$ over $\VspaceR$. 
The vector space $\VspaceR$ can be extended (or "complexified") to become a vector space  $\VspaceC$ over $\mathbb{C}$. Formally, the complexification $\VspaceC$ of $\VspaceR$ is defined as 
\be 
\VspaceC =  \mathbb{C} \otimes_\mathbb{R} \VspaceR
\ee
where the tensor product  above can just be considered as a tensor product between two vector spaces $\VspaceR$ and $\mathbb{C}$ over the field of real numbers (hence the notation $\otimes_\mathbb{R} $). 
Hence, an arbitrary vector in $\VspaceC$ is just given by an arbitrary linear combination of some tensor product $z \otimes_\mathbb{R} \bm{\mathcal V}$ where $z \in \mathbb{C}$ and $\bm{\mathcal V}$ is some vector in $\VspaceR$. For three arbitrary complex numbers $z_1$, $z_2$ and $z_3$ and two arbitrary vectors $\bm{\mathcal V}_1$ and $\bm{\mathcal V}_2$ inside $\VspaceR$, the composition law $z_1\left( \left(z_2+z_3\right) \otimes_\mathbb{R} \left(\bm{\mathcal V}_1 + \bm{\mathcal V}_2 \right)\right) = \left(z_1 z_2+z_1 z_3\right) \otimes_\mathbb{R} \left(\bm{\mathcal V}_1 + \bm{\mathcal V}_2\right)$ gives to $\VspaceC$ a structure of a $\mathbb{C}$-vector space. 
A simpler notation for an arbitrary vector $\bm{\mathcal V}$ of $\VspaceC$ is just 
$\bm{\mathcal V}  = l_1 \bm{\mathcal V}_1 + i \;l_2 \bm{\mathcal V}_2$ where $l_1$ and $l_2$ are real numbers and $ \bm{\mathcal V}_1$ and $\bm{\mathcal V}_2$ are elements of $\VspaceR$.
Furthermore, we can write $\VspaceC \equiv  \mathbb{C} \otimes_\mathbb{R} \VspaceR = \VspaceR \oplus i \;\VspaceR$. An endomorphism $\phi$  of $\VspaceR$ can also be "complexified" into an endomorphism $\phi_\mathbb{C}$  of $\VspaceC$ defining 
$\phi_\mathbb{C} \left(z \otimes_\mathbb{R} \bm{\mathcal V}\right) \equiv z \otimes_\mathbb{R} \phi \left(\bm{\mathcal V}\right) = Re(z) \phi \left(\bm{\mathcal V}\right) + i\; Im(z) \phi \left(\bm{\mathcal V}\right)$ and extending this by linearity. Now considering a representation $(\VspaceR,\rho)$  of the group $G$ on the vector-space $\VspaceR$ realized over the field $\mathbb{R}$, we obtain a new representation of  $G$, $(\VspaceC,\rho^\mathbb{C})$, on the vector space $\VspaceC$ by defining for all $g \in G$, $\rho^\mathbb{C}(g) = \left(\rho(g)\right)_\mathbb{C}$.
If an irreducible representation $(\Vspace_{\mathbb{R}}, \rho)$ is such that $(\VspaceC,\rho^\mathbb{C})$ is also irreducible, we say that it is \textbf{\textit{absolutely irreducible}}.  That is, an irreducible vector space remains irreducible under the "complexification" described above.\footnote{The notions introduced in this subsection and in the following one can be extended to vector spaces over an arbitrary fields $\mathbb{F}$ and and an arbitrary field extension of it.} This can be shown to be the case for representations of the symmetric group.\footnote{This is a consequence of the fact that irreducible representations of the symmetric group are defined on $\mathbb{Q}$ and stay irreducible on any extensions of $\mathbb{Q}$ of characteristic $0$ such as $\mathbb{R}$ and $\mathbb{C}$.}

\subsection{Schur's lemma and some corollaries} \label{AppA3}

Let $(\Vspace, \rho), (\Wspace, \rho')$ be two irreducible representations of the group $G$ and $\phi:\Vspace \to \Wspace$ be a \textit{linear} map that \textit{intertwines} with the group action, i.e. which is an intertwiner map as defined in the above subsection \ref{RepGen}
then 
\begin{enumerate}
\item $\phi$ is either 0 or an isomorphism
\item  In the particular case when $(\Vspace, \rho)$ = $(\Wspace, \rho')$, and the base field of the vector space $\Vspace$ is $\mathbb{C}$, then $\phi = \lambda \; \mathbb{1},$ with $\lambda \in \mathbb{C}$ and $\mathbb{1}$ the identity in $\mathbf{GL}(\Vspace)$. That is $\phi$ is a scalar multiple of the identity.
\end{enumerate}
The proof of the above lemma can be found in a variety of references (e.g. in \cite{Serre}), however, for the sake of clarity and of the understanding of the remaining of this subsection,  we found it useful to give it explicitly below.

\begin{proof}
The map $\phi = 0$ is a trivial intertwiner. Suppose $\phi \neq 0$ and consider the kernel of $\phi$, $Ker(\phi)$. If $\bm{\mathcal V} \in Ker(\phi)$ then $0 = \rho'_g(\phi(\bm{\mathcal V})\equiv 0) = \phi(\rho_g(\bm{\mathcal V})) \forall g \in G$. Hence $Ker(\phi) \subset \Vspace$ is an invariant subspace of the representation $(\Vspace, \rho)$. Since this representation is irreducible by hypothesis, any invariant subspace is either the \textit{null space} or the entire space $\Vspace$; the second case i.e. the case with $Ker(\phi) = \Vspace$ is excluded as $\phi \neq 0$. By a similar argument we can conclude that the image of $\phi$ is the entire space $\Wspace$, i.e. $Im(\phi) = \Wspace$. The properties $Ker(\phi) =0$ and $Im(\phi) = \Wspace$ imply that $\phi$ is a G-\textit{isomorphism} which proves the first part of the lemma. Consider now the case where $\Vspace = \Wspace$ and $\rho = \rho'$, and define the map $\phi' \equiv \phi - \lambda\; \mathbb{1}$ where $\lambda\in \mathbb{C}$ is a non zero eigenvalue of $\phi$ (which always exists since $\mathbb{C}$ is algebraically closed). Then $\phi'$ is an intertwiner and has a non-zero kernel. These properties can only be satisfied if $\phi' = 0$ which implies that $\phi = \lambda\; \mathbb{1}$. 
\end{proof}

The above Schur lemma also extends as follows to the case of an absolutely irreducible representation on the real field. Indeed, 
let $(\VspaceR,\rho)$ be an absolutely irreducible representation.  Then, any G-homomorphism $\phi: \Vspace_\mathbb{R} \longrightarrow \Vspace_\mathbb{R}$ is necessarily a scalar multiple of the identity {\it inside} $\mathbb{R}$, i.e.,
\be 
\phi = \lambda \;\mathbb{1} \quad \lambda \in \mathbb{R}
\ee
\begin{proof}
Let us consider the representation  $(\VspaceC,\rho^\mathbb{C})$ as defined above and the complexification $\phi_\mathbb{C}$ of the map $\phi$. The map $\phi_\mathbb{C}$ is a G-homomorphism of the irreducible representation $(\VspaceC,\rho^\mathbb{C})$, hence Schur's lemma states that it must be of the form $\phi_\mathbb{C} = \lambda \; \mathbb{1}$, with $\lambda$ some complex number.
Hence, for an arbitrary non-vanishing complex number $z$ and vector $\bm{\mathcal V}$ of $\VspaceR$ one has $\phi_\mathbb{C} \left( z \otimes_\mathbb{R} \bm{\mathcal V} \right) = \lambda \left( z \otimes_\mathbb{R} \bm{\mathcal V} \right)$ which reads also 
$\left( Re(z) + i \; Im(z)\right) \phi \left(\bm{\mathcal V}\right) =  \left( Re(\lambda) + i \; Im (\lambda \right) \left( Re(z) + i \; Im(z)\right) \bm{\mathcal V} = \left(Re(z)  Re(\lambda) - Im(z) Im(\lambda)\right)\bm{\mathcal V}  + i \left(Re(z) Im(\lambda) + Im(z) Re(\lambda)\right) \bm{\mathcal V}$ and implies $Im(\lambda)=0$ and also $\phi \left(\bm{\mathcal V}\right) = \lambda \bm{\mathcal V}$ ending the proof.
\end{proof}

The above lemmas lead to the following corollaries. First, Let $(\Vspace_1,\rho_1), (\Vspace_2,\rho_2)$ be equivalent irreducible representations over $\mathbb{C}$ or equivalent absolutely irreducible representations over $\mathbb{R}$, then the G-isomorphism that exists between them is unique up to scalar multiplication. Indeed, Let $\phi_i:\Vspace_1 \longrightarrow \Vspace_2 \quad i=1,2 $ be two G-isomorphisms. Then the map $\phi_2^{-1}\circ \phi_1: \Vspace_1 \longrightarrow \Vspace_1$ is a G-automorphism. It follows from Schur's lemma that one must  have $\phi_2^{-1}\circ \phi_1 = \lambda \; \mathbb{1}$ implying $\phi_1 = \lambda \;\phi_2, \lambda \in \mathbb{C} \;\; {\rm (or} \;\; \mathbb{R})$.

Secondly, let us consider an irreducible representation $(\Vspace, \rho)$ and let $\Wspace = \Vspace_1 \oplus \dots \oplus \Vspace_m$ where $\Vspace_i$ is isomorphic to  $\Vspace \forall i \in \{1,\cdots, m\}$, then the number of linearly independent G-homomorphisms $\psi_i: \Wspace \longrightarrow \Vspace$ is exactly $m$.
Indeed, let $\phi_i: \Vspace_i \longrightarrow \Vspace$ be non-zero G-isomorphisms (which are guaranteed to exist since $\Vspace_i \simeq \Vspace$). 
Let $\tilde{\pi}_{i}:\Wspace \longrightarrow \Vspace_i$, be the canonical projections on $\Vspace_i$, and we further define $\pi_{i}:\Wspace \longrightarrow \Wspace$ such that  $\forall \bm{\mathcal W} \;\in \;\Wspace$, $\pi_i\left(\bm{\mathcal W}\right) = \tilde{\pi}_i\left(\bm{\mathcal W}\right)$ (i.e. $\pi_i$ are the compound of $\tilde{\pi}_{i}$ by the inclusion map of $\Vspace_i$ inside $\Wspace$). The identity map inside $\Wspace$, $\mathbb{1}_\Wspace$, is just given by 
 $\mathbb{1}_\Wspace = \sum_i  \pi_i$. Now consider an arbitrary G-homomorphism $\psi: \Wspace \longrightarrow \Vspace$.
One has using the above decomposition of $\mathbb{1}_\Wspace$,
\be 
\psi \equiv \sum_{i=1}^m \psi_i\;\; 
\ee
where $\psi_i = \psi \circ \pi_i$ are maps between $\Wspace$ and $\Vspace$.  
The restriction of $\psi_i$ to $\Vspace_i$, defined to be $\tilde{\psi}_i$, is a G-homomorphism and hence, thanks to the preceding lemma, we have that 
$\tilde{\psi}_i = \lambda_i \phi_i$, with $\lambda_i$ some real number, which ends the proof.



\section{The group algebra of the symmetric group and its regular representations} \label{AppB}
\subsection{Definitions of the group algebra of $S_n$} \label{AppB1}
The so-called \textbf{\textit{group algebra}} (also called the \textbf{\textit{Frobenius algebra}} [see e.g. \cite{Littlewood} p43] or the \textbf{\textit{group ring}}) of $S_n$, that is usually noted as $\mathbb{R}[S_n]$ (for reasons that will also appear clear below) plays a crucial role in elucidating the link between tensor symmetries and representations of the symmetric group that we use in this work.

 A first definition of the group algebra is as follows: first, out of $S_n$ we can define a vector space $\Vspace_{S_n}$ over $\mathbb{R}$ by considering each different element $\sigma_i$ of $S_n$ as a base vector noted here $\mathbf{e}_{\sigma_i}$ (note that in the following we will sometimes use the slight abuse of notation consisting in identifying the permutation $\sigma_i$ with $\mathbf{e}_{\sigma_i}$ the corresponding basis element of $\mathbb{R}[S_n]$) and defining the elements of  $\Vspace_{S_n}$ (which will turn out to be identical\footnote{In the following we use sometimes the notation $\Vspace_{S_n}$ instead of $\mathbb{R}[S_n]$ when we want to stress that we only use the vector space structure of the group algebra, but we stress here again that the two sets $\Vspace_{S_n}$ and $\mathbb{R}[S_n]$ are just identical. In the literature, $\mathbb{R}[S_n]$ is usually called either the group algebra, the group ring, or the Frobenius algebra associated with $S_n$} to $\mathbb{R}[S_n]$) as just arbitrary linear combinations of the $\mathbf{e}_{\sigma_i}$ with real coefficients of the form 
 $\sum_{i=1}^{i=n!} l_{\sigma_i} \mathbf{e}_{\sigma_i}$ where $l_{\sigma_i}$ are $n!$ real numbers (and the product and sum appearing in this definition are so far formal - see below). The addition of two such vectors and multiplication by a real number, $l$, are just then defined in a natural way as 
\begin{eqnarray}
\left(\sum_{i=1}^{i=n!} k_{\sigma_i}\mathbf{e}_{\sigma_i}\right) + \left(\sum_{i=1}^{i=n!} l_{\sigma_i} \mathbf{e}_{\sigma_i}\right) = \sum_{i=1}^{i=n!} \left(k_{\sigma_i} +l_{\sigma_i}\right) \mathbf{e}_{\sigma_i} \\
l \left(\sum_{i=1}^{i=n!} k_{\sigma_i}\mathbf{e}_{\sigma_i}\right) = \sum_{i=1}^{i=n!} l k_{\sigma_i}\mathbf{e}_{\sigma_i}.
\end{eqnarray}
Using then the composition law of $S_n$ as a internal composition law on $\Vspace_{S_n}$ as (note that below and in the following we note this law with the $\times$ symbol, but we will also later omit this symbol in order to alleviate notations) 
\begin{eqnarray} \label{RepDefVSn}
\left(\sum_{j=1}^{j=n!} k_{\sigma_j} \mathbf{e}_{\sigma_j}\right) \times\left(\sum_{i=1}^{i=n!} l_{\sigma_i}\mathbf{e}_{\sigma_i}\right) &=& \sum_{j=1}^{j=n!} \sum_{i=1}^{i=n!}\left(k_{\sigma_j} l_{\sigma_i}\right) \mathbf{e}_{\sigma_j \sigma_i  } 
\end{eqnarray}
gives to $\Vspace_{S_n}$ the structure of an algebra. 
We note that $S_n$ is obviously included into its group algebra  $\mathbb{R}[S_n]$, and, if we consider the element of the group algebra in the left parenthesis of the left-hand side of equation (\ref{RepDefVSn}) just to be given by an element of $S_n$ (i.e. setting all but one $k_{\sigma_j}$ to zero), we can interpret (\ref{RepDefVSn}) as defining a representation of the permutation group $S_n$ on the vector space  $\Vspace_{S_n}$ where the action of a group element on an element of the group algebra (considered as a vector space) is simply given here by the left product of the group element by the vector. This representation is called the \textbf{\textit{(left) regular representation}} of $S_n$. The same action (now extended to a given element of the group algebra in the left parenthesis of the left hand side of equality (\ref{RepDefVSn}) - i.e. allowing for arbitrary $k_{\sigma_j}$) clearly also provides a representation of the full group algebra acting on itself (as a vector space) by left multiplication.  In general every representation of the group algebra also contains one of the group itself (because the group is contained into the group algebra) and conversely for any representation of the group there exists a unique representation of the group algebra in which it is contained (which is just obtained by considering formal linear combinations of matrices associated with the considered group representation.  One can further show that if a representation is irreducible or reducible  as a representation of the group algebra, then it has the same property as a representation of the group.

There is another definition of the group algebra which is handy for us. $\mathbb{R}[S_n]$ can just be defined as the set of real functions on the symmetric group $S_n$. 
Indeed, because $S_n$ is finite, any such function $f$ is fully defined by $n!$ real numbers $f(\sigma_i)$ where $\sigma_i$ runs over all the elements of $S_n$. This provides a one to one map between the set of real functions on $S_n$ and $\mathbb{R}[S_n]$,
 which sends a given function $f$ to the group algebra element $\sum_{i=1}^{i=n!} f(\sigma_i) \mathbf{e}_{\sigma_i}$. Using this definition, there is a natural operation that can be defined on the group algebra and which will play a role in the following: it is called the \textbf{\textit{"$*$" involution map}}: it sends the group algebra element 
 $a = \sum_{i=1}^{i=n!} f(\sigma_i) \mathbf{e}_{\sigma_i}$ to $a^* = \sum_{i=1}^{i=n!} f(\sigma_i) \mathbf{e}_{\sigma_i^{-1}}$. 
\subsection{Young diagrams, Young tableaux and Young symmetrizers} \label{AppB2}
As it will be recalled later, the regular representation of the group algebra $\mathbb{R}[S_n]$ is fully reducible and its decomposition into irreducibles allows to fully classify tensor symmetries and operate a similar decomposition of a given space-time tensor. This decomposition in practice relies on the use of special elements of the group algebra which are called Young symmetrizers and which will be defined below. To do so, we first need to introduce some basic notions about partitions, Young tableaux and diagrams. These notions, as well as some properties to be given below, will also be used in this article when dealing with plethysms and symmetric functions as well as to derive our main results.   

We first recall that, given a positive integer $n$, a \textbf{\textit{partition}} $\lambda$ of $n$, denoted by $\lambda \vdash n$, is a sequence of positive integers, $(\lambda_1, \lambda_2, \dots \lambda_r)$, such that $\lambda_1 \geq \lambda_2 \geq \dots \geq \lambda_r$ and $\lambda_1 + \lambda_2 \dots +\lambda_r = n$.

A \textbf{\textit{Young diagram}} is a finite collection of boxes arranged in left-justified rows, with the row sizes weakly decreasing (this being the so-called "english" notation for Young diagram, while the "french" convention is upside down with respect to the "english" one).\footnote{See e.g. Refs \cite{Fulton,Tung:1985na} for nice reviews on Young diagrams and tableaux.} 
 Each Young diagram corresponds uniquely to a partition and will sometimes be noted as the corresponding partition $\lambda = (\lambda_1, \lambda_2, \dots \lambda_r)$. The Young diagram associated to a given $\lambda$, is the one that has $r$ rows (using the notation of the above definition) and $\lambda_i$ boxes on $i$'th row. For example, the partition (3,2,2) corresponds to the following Young diagram.
\\\begin{center}
\ydiagram{3,2,2}
\end{center}
With the above notation, a \textbf{\textit{Young diagram with one line of $\bm{m}$ boxes}} with just be denoted (when it is clear from the context that we consider a Young diagram and not an integer) as $\bm{(m)}$.  We will sometimes use a simplifying notation to designate \textbf{\textit{Young diagrams with several lines each of the same length $\bm{m}$}}. Such a diagram with $p$ such lines will be denotes as  $\bm{(m)^p}$.


Given a Young diagram corresponding to a partition $(\lambda_1, \dots \lambda_r)$, the \textbf{\textit{conjugate}} or \textbf{\textit{transpose}} of this Young diagram is given by the reflection of the original diagram along its main diagonal, denoted $(\lambda_1 \dots \lambda_r)^t$ where the superscript $"t"$ stands for transpose. The transpose of our above example is simply $(3,2,2)^t$ and corresponds to the following Young diagram
\\\begin{center}
\ydiagram{3,3,1}
\end{center}

Given $n$ ordered labels such as $a\{n\} := \{a_1,\dots ,a_n\}$, a \textbf{\textit{Young tableau}} $ \lambda_{k} \equiv \lambda_{a\{n\}}$ is a filling of the Young diagram $\lambda$ with labels $a_i$, one in each box and such that a given label $a_i$ can be used several times to fill a box. In the notation $ \lambda_{k} $,  the subscript $"k"$ labels different Young tableaux derived from the same Young diagram $\lambda$ but different fillings.  
For example given the labels, $\{1,2,3\}$ and a partition $(2,1)$, admissible Young tableaux are 

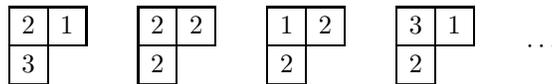
\ytableausetup 
{mathmode, boxsize=0.5cm, centertableaux}
\begin{center}
\ytableaushort {21,3} \qquad \ytableaushort {22,2}\qquad \ytableaushort {12,2} \qquad \ytableaushort {31,2} \qquad $\cdots$
\end{center}

A \textbf{\textit{standard Young tableau}} is a Young tableau filled such that the order is preserved strictly along the rows (from left to right) and the columns (from top to bottom) and that each label occurs at most once. For example given the labels, $\{1,2,3\}$ (ordered using the natural order on $\mathbb{N}$)  and a partition $(2,1)$ all the associated standard Young tableaux are
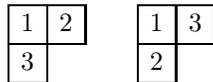
\ytableausetup 
{mathmode, boxsize=0.5cm, centertableaux}
\begin{center}
\ytableaushort {12,3} \qquad \ytableaushort {13,2}
\end{center}
For a given  young diagram $\lambda$ (and a set of labels) we denote by $\bm{\STl}$ the set of all standard Young tableaux built from $\lambda$.

A \textbf{\textit{semi-standard Young tableau}} is a Young tableau filled such that the order is preserved strictly along columns but only weakly along the rows. E.g. all the semi-standard Young tableaux built from the 
partition $(2,1)$ and with entries in the labels, $\{1,2,3\}$ are 
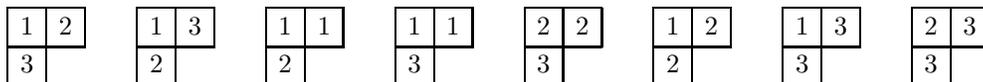
\ytableausetup 
{mathmode, boxsize=0.5cm, centertableaux}
\begin{center}
\ytableaushort {12,3} \qquad \ytableaushort {13,2} \qquad \ytableaushort {11,2} \qquad \ytableaushort {11,3} \qquad \ytableaushort {22,3} \qquad \ytableaushort {12,2} \qquad \ytableaushort {13,3} \qquad \ytableaushort {23,3}
\end{center}
Considering a given semi-standard Young tableau filled with integers $\{1,2,3, \cdots\}$, this tableau is said to have \textbf{\textit{type}} $\bm{\alpha=(\alpha_1, \alpha_2, \cdots)}$ if the number of occurrence of the integer $i$ is equal to $\alpha_i$.

Consider now a given Young diagram $\lambda$ and some Young tableau $\lambda_k \equiv \lambda_{a\{n\}} $ obtained from it and an ordered list of label $a\{n\}$. There is a natural action of $S_n$ on $\lambda_k$  defined (with obvious notations) by $\sigma \left( \lambda_{a\{n\}} \right) = \lambda_{\{a_{\sigma(1)}, \cdots, a_{\sigma(n)}\}}$, where $\sigma$ is an element of $S_n$ and this action is simply defined by the permutation of the labels of the boxes of the tableau using $\sigma$. Furthermore, let $\bm{\Rgroup \subset S_n$}\textbf{, be the \textit{row group}} associated to the Young tableau $\lambda_k$ 
 defined as the subgroup of $S_n$ which leaves the set of elements in each row invariant (i.e. it can only change the order of the elements in a given row).  Similarly, let $\bm{\Cgroup \subset S_n}${\textbf{, be the \textit{column group}}, i.e. the subgroup of $S_n$ that preserves the set of elements of each column.
Out of $\Rgroup$ and $\Cgroup$ associated to a given tableau, we can define the following elements belonging to the group algebra $\mathbb{R}[S_n]$
\be 
\Rsym = \sum_{\sigma \in \Rgroup} \bm{e}_\sigma \qquad \Csym = \sum_{\sigma \in \Cgroup} \sign(\sigma) \bm{e}_\sigma.
\ee
For a given tableau $\lambda_k$, one can further define a ("symmetric")  \textbf{\textit{Young  symmetrizer}} as the element 
of the group algebra given by 
\be \label{youngsymdef}
\Youngsym \equiv \Rsym \times \Csym.
\ee
Due to the order of the multiplication above, this element is such that it is automatically symmetric in the labels corresponding to the rows of the Young tableau, which justifies the upper "sym" in our notation. This corresponds to the definition of Young symmetrizers used in most references, however, another set of Young symmetrizers can be used which are this time explicitly antisymmetric in the labels corresponding to columns. These "antisymmetric" Young symmetrizers are defined by 
\be \label{youngantidef}
\Younganti \equiv \Csym \times \Rsym.
\ee
By the action of the $*$ involution on a given Young symmetrizer, one has obviously
$(\Rsym)^*=\Rsym$, $(\Csym)^*=\Csym$ and $(V_1\times V_2)^*=(V_2)^*\times (V_1)^* \;\;\forall V_{1,2}\in \mathbb{R}[S_n]$, and thus 
\be \label{involutionYoung}
\left(\Youngsym\right)^* = \Younganti \qquad {\rm and} \qquad \left(\Younganti\right)^* = \Youngsym.
\ee

A Young symmetrizer is what is called an \textbf{\textit{essentially idempotent}} of the group ring, i.e. an element $\bm{y}$ of $\mathbb{R}[S_n]$ which verifies $\bm{y}^2 = l \bm{y}$ where $l $ is a real number. It is such that $\tilde{\bm{y}} = \bm{y}/l$ is an \textbf{\textit{idempotent}}, i.e. verifies $\tilde{\bm{y}}^2 = \tilde{\bm{y}}$ (see e.g. \cite{Boerner} p 103). Accordingly, we call $\Youngsymantitilde$ the idempotent associated to the essentially idempotent $\Youngsymanti$.

\subsection{Ideals of $\mathbb{R}[S_n]$ and the full reduction of the regular representation} \label{AppB3}

We first recall that a \textbf{\textit{left (respectively right) ideal}} inside a ring $X$ (for us, the relevant ring to consider here is the group ring which is the same as the group algebra $\mathbb{R}[S_n]$) is a subset $I$ of the ring such that i/ it is a subgroup of the ring for the addition law (i.e. the internal law for which $X$ is a group) and ii/ for any $\bm{x}$ belonging to $X$ and any $\bm{i}$ belonging to $I$ the product $\bm{x} \times \bm{i}$  (respectively $\bm{i} \times \bm{x}$, where $\times$ denotes the internal product of the ring) belongs to $I$. A \textbf{\textit{two-sided ideal}} is an ideal which is at the same time a left and a right ideal. An ideal (left, right or two-sided) is said to be \textbf{\textit{minimal}} when it contains no other ideal (of the same kind) besides the trivial ideals, i.e. itself and the null ideal. 

There is an easy way to construct ideals: consider a given element $\bm{g}$ of the group algebra and then the set of all the right product of an arbitrary element $\bm{x}$ of $\mathbb{R}[S_n]$ with $\bm{g}$, i.e. $\{ \bm{x} \times \bm{g}, \; \bm{x}\in R\}$. This set can be noted $\mathbb{R}[S_n] \bm{g}$ and is obviously a left ideal. It is said to be \textbf{\textit{generated}} by $\bm{g}$. Similarly, the set of the solutions $\bm{x}$ of the equation $\bm{x} \times \bm{g} = 0 $ (where $0$ is the null vector and $\bm{g}$ is a given element of the ring) also constitutes a left ideal. One can show that every ideal can be generated in both ways. In particular, it can be shown that every left ideal $I$ of $\mathbb{R}[S_n]$ contains an element $\bm{e}$ such that $\bm{e}$ is an idempotent (i.e. verifies $\bm{e}^2 =\bm{e}$) and generates $I$ (i.e. $I= \mathbb{R}[S_n]~\bm{e}$) (see \cite{Boerner} p 58-59). The element $\bm{e}$ is called a \textbf{\textit{generating idempotent}} of $I$. If the ideal $I$ is minimal, then every generating idempotent of $I$ is 
\textbf{\textit{primitive}}.\footnote{See \cite{Boerner} p 60 for a proof. A primitive idempotent $\bm{g}$ is an idempotent which cannot be written as a sum of two different (and non vanishing) idempotent $\bm{g}'$ and $\bm{g}"$ which would in addition be orthogonal, i.e. verify $\bm{g}'\times \bm{g}" = 0$ and $\bm{g}"\times \bm{g}'=0$.} And conversely any primitive idempotent $\bm{e}$ generates a minimal left ideal $
\mathbb{R}[S_n]~\bm{e}$. Note also that the intersection of a finite number of left (respectively right or two-sided) ideals is an ideal of the same type.

The ideals of the group ring $\mathbb{R}[S_n]$ play a crucial role in the decomposition of the regular representation defined above. Indeed, the \textbf{\textit{invariant subspaces}} of this representation can be shown to be left ideals while \textbf{\textit{irreducible spaces}} are minimal such ideals. The whole group ring is a direct sum of minimal left ideals. This decomposition is unique except for the order and up to equivalence between minimal left ideals.\footnote{Two ideals are said to be equivalent in this context if there is a linear map between them which is compatible with left multiplication. I.e. if for any element $\bm{i}$ of one ideal mapped to the element $\bm{i'}$ of the other ideal, then $\bm{x} \times \bm{i}$ is mapped to $ \bm{x} \times \bm{i'}$ for any $\bm{x}$ belonging to the group ring $\mathbb{R}[S_n]$. One can show that any such map is a right multiplication.}
Indeed, the group algebra can be decomposed in the following way
\be \label{Decirredleft}
\mathbb{R}[S_n] = \bigoplus_{\lambda \vdash n } I_{\lambda}
\ee
where the sum runs over all the partitions $\lambda$ of $n$ (which are in one to one correspondence with Young diagrams, and which we note here as $\lambda \vdash n$) and $I_\lambda$ is a two sided ideal in one to one correspondence with a given Young diagram $\lambda$. Each ideal $I_\lambda$ can be further decomposed in a direct sum of left ideals $L_{\lambda_k}$ as follows 
\be \label{IlambdabyLlambdak}
I_{\lambda} =\bigoplus_{\lambda_k \in ST_{\lambda}} L_{\lambda_k}, \quad
L_{\lambda_k} = \mathbb{R}[S_n]~\Youngsymanti ,
\ee
where the sum now runs over the set of all standard Young tableaux $ST_{\lambda}$ which can be built out of a given partition $\lambda$, and $\Youngsymanti$ is the Young symmetrizer constructed from the standard Young tableau $\lambda_k$ in the antisymmetric or symmetric presentation\footnote{Note that the same presentation should be chosen for all the Young symmetrizers entering into the decomposition}. Each Young symmetrizer $\Youngsymanti$ is a generating essentially idempotent of the ideal  $L_{\lambda_k} \equiv \mathbb{R}[S_n]~\Youngsymanti$, and this ideal is minimal. Note further that while Young symmetrizers associated to different Young diagrams are mutually orthogonal, the young symmetrizers associated to the same Young diagram but different standard tableaux are not mutually orthogonal. Putting all together we then have the full decomposition of the regular representation into irreducible representations of the symmetric group (or of the group algebra) each defined by the left action 
(\ref{RepDefVSn}) on the irreducible spaces $\mathbb{R}[S_n]~\Youngsymanti$ given by 
\be \label{Decirredleftbis}
\mathbb{R}[S_n] = \bigoplus_{\lambda \vdash n } \bigoplus_{\lambda_k \in ST_{\lambda}} \mathbb{R}[S_n]~\Youngsymanti.
\ee
 One can show that for a given Young diagram $\lambda$, the dimension of  $\mathbb{R}[S_n]~\Youngsymanti$ as a vector space over $\mathbb{R}$ is the same as the number of different standard Young tableaux that can be built out of $\lambda$ which also gives the number of isomorphic (but inequivalent) left ideals $\mathbb{R}[S_n]~\Youngsymanti$ entering into the decomposition of the group ring.

A similar decomposition can be obtained using right ideals (i.e. decomposing the two sided ideals $I_{\lambda}$ into direct sums of right ideals). Indeed, one has analogously to (\ref{Decirredleft})
\be \label{Decirredright}
\mathbb{R}[S_n] = \bigoplus_{\lambda } \bigoplus_{\lambda_k \in ST_{\lambda}} \Youngantisym ~\mathbb{R}[S_n].
\ee
This decomposition can be obtained by applying the $*$ involution map to (\ref{Decirredleftbis}) and using (\ref{involutionYoung}). Indeed, the $*$ involution map  maps idempotents to idempotents, direct sums of left (right) ideals to direct sums of right (left) ideals and minimal ideals to minimal ideals.

 If one multiplies to the right the above equality (\ref{Decirredleft}) by some arbitrary element $\bm{g}$ of the group algebra, one obtains a decomposition of the left ideal $L \equiv \mathbb{R}[S_n]~\bm{g}$ generated by $\bm{g}$ (i.e. of any left ideal of $\mathbb{R}[S_n]$, since any such ideal has a generating idempotent) as 
\be \label{decideal}
L \equiv \mathbb{R}[S_n]~\bm{g} = \sum_{\lambda } \sum_{\lambda_k \in ST_{\lambda}} \mathbb{R}[S_n]~\Youngsymanti\times\bm{g} 
\ee
however, the sum on the right hand side above is no longer direct. Still, this can be used to obtain a decomposition of $L$ into a direct sum of minimal left ideals. This decomposition can be obtained from the above (\ref{decideal}) just by removing there a sufficient number of the minimal ideals $\mathbb{R}[S_n]~\Youngsymanti\times\bm{g} $ (one can indeed show that these ideals are minimal). Fiedler \cite{Fiedler1,FiedlerHab} gives an algorithm to carry out this decomposition. It uses the fact that the minimal left ideals $\mathbb{R}[S_n]~\bm{y}_{\lambda}^{anti/sym}\times\bm{g} $ and $\mathbb{R}[S_n]~\bm{y}_{\mu}^{anti/sym}\times\bm{g} $ are in direct sum if $\mu$ and $\lambda$ are standard tableaux each corresponding to different partitions of $n$ (i.e. to different Young diagrams). Hence, it is only the ideals of the form $\mathbb{R}[S_n]~\bm{y}_{\lambda_k}^{anti/sym} \times \bm{g} $ with the $\lambda_k$ corresponding to the same young diagram (but to different standard tableaux) which are not necessarily in direct sum in the expression above, depending on the considered element $\bm{g}$ of the group algebra. This means in particular that it is only whenever, for a given fixed young diagram, there is more than one standard young tableau entering in the sum (\ref{IlambdabyLlambdak}) that one has possibly to remove corresponding terms in the sum (\ref{decideal}). 
The algorithm devised by Fiedler goes then as follows: for a given partition $\lambda \vdash n$, keep the first non vanishing ideal $\mathbb{R}[S_n]~\bm{y}_{\lambda_k}^{anti/sym} \times \bm{g} $ entering in the sum on the right hand side of (\ref{decideal}), and call it $L_1$. The next non vanishing ideal in the sum can be shown to be either contained in $L_1$ or not. In the latter case, call it $L_2$. $L_1$ and $L_2$ are in direct sum and one can call $\tilde{L}_2$ their direct sum. The algorithm continues then in the same way by replacing $L_1$ by $\tilde{L}_2$ until one has exhausted all the left ideals on the right hand side of (\ref{decideal}) corresponding to the same partition $\lambda$ of $n$, the ideals corresponding to different partitions being automatically in direct sum.
The algorithm given in \cite{Fiedler1,FiedlerHab} enables also to construct a generating idempotent $\bm{e}$ of the left ideal $L$ and its decompositions into primitive pairwise orthogonal idempotents $\bm{e}_i$ such that $\bm{e} = \bm{e}_1 + \cdots + \bm{e}_k$ and that the ideals $L_k$ generated by each $\bm{e}_k$ are in direct sum and sum up to $L$.

A similar decomposition can also be achieved for right ideals. Indeed, one way to proceed considering an arbitrary right ideal $R$ is to decompose the left ideal $L = R^*$ as shown above and then come back to $R$ by acting again on the decomposition of $L$ into direct sum of minimal left ideal with the $*$ involution map. 

This decomposition for left or right ideals can be used to decompose any space-time tensor into components with given symmetries, as we now explain.

\label{Idealreducsubsection}

\section{Tensors, Tensor symmetries and the group algebra $\mathbb{R}[S_n]$} \label{AppC}
\subsection{From abstract and space-time tensors to the group algebra of $S_n$} \label{AppC1}

The group algebra can be defined in a way that makes its links to tensor more explicit, namely as follows: consider first a chosen ordered list of labels $a\{n\}=\{a_1,\cdots,a_n\}$  which do not yet have any meaning as space-time indices. Then we can identify the identity inside $S_n$ with this list and further any permutation $\sigma$ of $S_n$ with the ordered list $\{a_{\sigma(1)} a_{\sigma(2)} \cdots a_{\sigma(n)}\}$. We can then think as these lists to be carried by some object  ${\mathbf T}$, such that the list $\{a_{\sigma(1)} a_{\sigma(2)} \cdots a_{\sigma(n)}\}$ is just identified with the list carried by $\mathbf{T}$ that we note as $T^{a_{\sigma(1)}a_{\sigma(2)} \cdots a_{\sigma(n)}}$. Here the object ${\mathbf T}$ has not to be considered as a space-time tensor, but just as an "abstract tensor" in the spirit of 
Ref.~\cite{Abstracttensor}, i.e. just an object of undetermined nature indexed by a string of character.
With such an identification, the group algebra is isomorphic (as an algebra) to the set of all (formal) linear combinations with real coefficients of all the indexed objects $T^{a_{\sigma(1)}a_{\sigma(2)} \cdots a_{\sigma(n)}}$ when $\sigma$ varies over $S_n$ , which can also be defined as the set $\bm{M}_{\bm{T}}= \spann \{ \sigma(\mathbf{T}) | \sigma \in S_n \}$ where "\textbf{\textit{span}}" means that we take the set of all possible linear combinations with real coefficients, and $\mathbf{T}$ and $\sigma(\mathbf{T})$ are defined (formally, since here, so far we deal only with abstract tensors $\bm{T}$ as opposed to space-time tensors noted with curly characters such as $\bm{\T}$) as in (\ref{actiontensor}).
As defined, $\bm{M}_{\bm{T}}$ is just the same as the group algebra $\mathbb{R}[S_n]$. So defined, we also have an action of $S_n$ on this set defined by $\rho \in S_n : T^{a\{\sigma(p)\}} \rightarrow T^{a\{\rho \times \sigma(p)\}}$ which is also identical to the action discussed above of $S_n$ on a given Young tableau (which can be considered as an abstract tensor in the sense above).

A set like $\bm{M}_{\bm{T}}$ also makes sense if we now consider a given space-time tensor $\bm{\T}$. More precisely, given a space-time tensor $\bm{\T}$ we define the set 
\be \label{SetMT}
\bm{M}
_{\bm{\T}} \equiv \spann \{ \sigma(\bm{\T}) | \sigma \in S_n \} = \mathbb{R}[S_n] {\bm{\T}}
\ee
and $\sigma(\bm{\T})$ defined as in (\ref{actiontensor}). This set $\bm{M}_{\bm{\T}}$
is a vector subspace (linear subspace) of the vector space of tensors $\Tn$. It is also the orbit of $\bm{\T}$ under the left action of the group algebra $\mathbb{R}[S_n]$ of $S_n$, where the left action is defined through (\ref{actiontensor}) and its natural linear extension. As such it provides a representation of 
$\mathbb{R}[S_n]$ (see e.g. Ref. \cite{Serre} p.5). 
The map $\psi : \mathbb{R}[S_n] \rightarrow \bm{M}_{\bm{\T}}$, which sends a permutation $\sigma$ to $\sigma\left(\bm{\T}\right)$ and is extended by linearity to linear combinations, is trivially an $S_n$-homomorphism.

Note that the $\bm{M}_{\bm{T}}$ and $\bm{M}_{\bm{\T}}$ should not be confused. On one hand, $\bm{M}_{\bm{T}}$ is equivalent to the group algebra $\mathbb{R}[S_n]$. On the other hand, as a vector space on $\mathbb{R}$, the group algebra has dimension $n!$, $\Tn$ has dimension $D^n$, and thus in general $\bm{M}_{\bm{\T}}$ is not isomorphic (as a vector space) to  $\mathbb{R}[S_n]$. (Consider e.g. the case with $n! > D^n$, and also the cases where the space-time tensor $\bm{\T}$ appearing in the definition of $\bm{M}_{\bm{\T}}$ obeys some relations of the type (\ref{defsym}) - in this case the images of all the $\sigma$ by $\psi$ are not all independent - moreover, in general a space-time tensor is not the same as an abstract tensor, since the former has real components while the latter is just a list of labels.)  However, the fact that the map $\psi$ defined in the previous paragraph is an $S_n$-homomorphism allows one to decompose $\bm{M}_{\bm{\T}}$  into irreducible representations of $S_n$, which corresponds, as we will see below, to subspaces of so-called \textbf{\textit{tensor symmetry classes}}. 
Indeed, according to the last lemma of appendix \ref{AppA1}, and using the decomposition of the group Algebra (\ref{Decirredleftbis}), one sees that for any tensor $\bm{\T}$ such that $\Youngsymanti \bm{\T}$ does not vanish, the image of the left ideal $\mathbb{R}[S_n] \Youngsymanti$ by $\psi$, that we can note $\mathbb{R}[S_n]\Youngsymanti \bm{\T}$, provides an irreducible representation of $S_n$ (see e.g. Ref. \cite{Tung:1985na} page 73). 

Given a tensor $\bm{\T}$ there is another useful way to build a corresponding subset inside the group algebra. Consider indeed first an arbitrary set of $n$ one forms  $\vcovect \equiv \{\vcovect^1, \dots , \vcovect^n\}$, where we allow repetition and linear dependence among the forms $\vcovect^1, \dots , \vcovect^n$. Out of $\bm{\T}$ and the n-uplet $\vcovect$ we can easily build a function on $S_n$ as follows: to each permutation $\sigma$ we associate the real number $\T_{\vcovect}\left(\sigma\right) \equiv \T^{b_1 \cdots b_n} \vcovect_{b_1}^{\sigma(1)} \cdots \vcovect_{b_n}^{\sigma(n)}$.  
We identify then this function with the element of the group algebra given by $\sum_{i=1}^{i=n!} \T_{\vcovect}\left( \sigma_i\right)\mathbf{e}_{\sigma_i}$ which we denote as $\T_{\vcovect}$. Note that if we choose a given ordered list of space-time indices values $a\{p\}$ (i.e. where each of the $a_i$ is an integer between 1 and D), and the set $\vcovect^i $ to be equal to the dual base covector $\bcovect^{a_{i}}$ (for which we have $\bcovect^{a_{i}}_{\mu}= \delta^{a_{i}}_{\mu}$) we have that $\T_{\vcovect}\left(\sigma\right) = \T^{b_1 \cdots b_n} \vcovect_{b_1}^{\sigma(1)} \cdots \vcovect_{b_n}^{\sigma(n)} = \T^{a_{\sigma(1)} \cdots a_{\sigma(n)}} = \left(\sigma\left(\mathbf{\T}\right)\right)^{a_1\cdots a_n}$ (see (\ref{actiontensor})). It is quite obvious that  
two tensors $\bm{\T}$ and $\bm{\S}$ (of the same valence) are equal iff they verify $\bm{\T}_{\vcovect} = \bm{\S}_{\vcovect}$ for any $\vcovect$ \cite{FiedlerHab}. Another result of interest for us is the following formula (demonstrated in \cite{FiedlerHab}) valid for any element $a$ of the group algebra acting on a tensor $\bm{\T}$
\be \label{ATstar}
\left(a \T\right)_{\vcovect}= \T_{\vcovect}\times a^*.
\ee

\subsection{Tensor symmetries and tensor symmetry classes} \label{symclass} \label{AppC2}

Our main tool in this work is the characterization of tensor symmetries and the purpose of this subsection is to explain how such symmetries are connected to the decomposition of the group algebra into irreducible representations of the symmetric group. To this aim, consider a tensor $\bm{\T}$ fulfilling one or several identities of the form (\ref{defsym}) which can also be encoded as 
\begin{eqnarray} \label{Tu}
\bm{u}_j \left( \bm{\T}\right) = 0, \;\; {\rm for}\;\; j=1,\cdots, m
\end{eqnarray}
where the $\bm{u}_j$ are $m$ elements of the group algebra $\mathbb{R}[S_n]$ and where the action of $\bm{u}_j$ on $\bm{\T}$ is defined by (\ref{actiontensor}) and linearity. Alternatively, these identities encoding the symmetries of $\bm{\T}$ can be written as 
\begin{eqnarray} \label{Tv}
\bm{v}_j \left( \bm{\T}\right) =\bm{\T}, \;\; {\rm for}\;\; j=1,\cdots, m
\end{eqnarray}
where $\bm{v}_j$ are $m$ elements of the group algebra $\mathbb{R}[S_n]$. The group algebra elements $\bm{u}_j$ or $\bm{v}_j$ characterize the symmetry of the considered tensor $\bm{\T}$ and for future reference we define the set $\Uset$ and $\Vset$ as the sets of group algebra elements $\Uset= \{\bm{u}_1, \cdots ,\bm{u}_m\}$ and $\Vset= \{\bm{v}_1, \cdots ,\bm{v}_m\}$ respectively. It is then natural to define the set of tensors of $\Tn$ which are invariant under the same symmetries as $\T$, i.e this set can be defined as $\Tn_{\Vset}$ by 
\ba
\Tn_{\Vset} = \{\bm{\X} \in \Tn, \forall \bm{v}_i \in \Vset \quad \bm{v}_i \left(\bm{\X}\right) = \bm{\X} \}.
\ea
It is easy to see that $\Tn_{\Vset}$ is a linear subspace of $\Tn$ that is invariant under the action of the bisymmetric transformations (\ref{actionGL}), simply because any such transformation commutes with any element of $\mathbb{R}[S_n]$ and in particular with the elements of $\Vset$. Now consider a subspace $\Wspace$ of $\Tn$ invariant under the bisymmetric transformations. It can be shown \cite{Boerner} that (i) this subspace 
 possesses generating idempotents $\bm{e} \in \mathbb{R}[S_n]$, in the sense that 
\ba \label{genidemvect}
\forall \bm{\X} \in \Tn, \quad \bm{e}(\bm{\X})\in \Wspace \quad {\rm and }\quad \forall \bm{\X} \in  \Wspace, \quad  \bm{e}(\bm{\X}) = \bm{\X}. 
\ea
The so-defined indempotent $\bm{e}$ is also  generating idempotent of the group algebra $\mathbb{R}[S_n]$.\footnote{A practical way to construct such an idempotent is to consider the projection on $\Wspace$ along any other linear space with which  $\Wspace$ is in direct sum to the full tensor space $\Tn$. This projection is then given by the product of some element of $\mathbb{R}[S_n]$ and the idempotent that one is looking for.}
One can also show (ii) that $\Wspace$ is such that the set $R$ of elements $\bm{r}$ of $\mathbb{R}[S_n]$, defined by 
\ba
R = \{ \bm{r}\in \mathbb{R}[S_n], \quad {\rm such \; that} \quad \forall \bm{\X} \in \Tn\, \quad \bm{r}(\bm{\X}) \in \Wspace\},
\ea
is a right ideal which is generated by any idempotent of $\Wspace$ as defined above. 
When $R$ is built as above from the invariant subspace $\Tn_{\Vset}$ corresponding to a given set of symmetries $\Vset$ of some given tensor $\bm{\T}$, we will denote it as $R_{v}$.
Given a right ideal $R$ inside $\mathbb{R}[S_n]$, we can define the set $\Tn_{R}$ as 
\ba
\Tn_R = \{ \bm{r}(\bm{\X}), \quad {\rm for\;  all} \quad \bm{r} \in R \quad {\rm and } \quad   \bm{\X} \in \Tn\}.
\ea 
This set is called a \textbf{\textit{tensor symmetry class}} and is invariant under the bisymmetric transformations. When considering such a set built from an ideal $R_{\Vset}$, we will note the associated tensor symmetry class as $\bm{V}_{\Vset}$ instead of the heavier $\Tn_{R_{\Vset}}$. We will use an alternative notation when the symmetries considered in $\Vset$ is of the Plethysm type defined in section \ref{Plethysms}. E.g. when we consider the plethysm $\plethysm$, we will denote by 
$R_{\plethysm}$ and $\bm{V}_{\plethysm}$ the corresponding right ideal and symmetry class. Similarly, when considering the right ideal $R_{\lambda_k}\equiv \Youngsymanti \times \mathbb{R}[S_n]$ generated by some specific Young symmetrizer $\Youngsymanti$ associated with a given Young tableau $\lambda_k$, we will denote as $\bm{V}_{\lambda_k}$ the corresponding symmetry class.

What matters for us is that $R_{\Vset}$ and the associated symmetry class $\Tn_{R_{\Vset}}$ can be simultaneously reduced respectively as representations of $S_n$ and of the bisymmetric (or $GL_D$) transformations. 
One way to proceed is to decompose the right ideal $R_{\Vset}$ into a direct sum of minimal ideals $R_{\bm{e}_k}$, $R_{\Vset}= \oplus_{\bm{e}_k} R_{\bm{e}_k}$, each generated by the idempotent $\bm{e}_k$ (e.g. using the method explained at the end of \ref{Idealreducsubsection}), then $\Tn_{R_{\Vset}}$ is decomposed into a direct sum of vector subspace each generated by one $\bm{e}_k$ in the sense of (\ref{genidemvect}), each if these subspace being an irreducible vector space under the action of the bisymmetric transformations. 

Alternatively, in order to use left ideals which are more easily connected with irreducible representations of the group algebra, one can use the following correspondence between tensor symmetry classes and left ideal essentially due to Weyl (see e.g. \cite{Weyl}) and nicely explained in \cite{Boerner,Fiedler1,FiedlerHab}. For a given tensor $\bm{\T}$ and its associated symmetry set $\Uset$ defined as above, as well as a given n-uplet 
 $\vcovect$ defined as above, all 
$\T_{\vcovect}$ belong to the left ideal $L$ defined as the intersection of the ideals $L_j$ defined by the set of elements of the group algebra which are annihilated by the group algebra elements $\bm{u}_j^*$. I.e. $L_j = \{ \bm{u} \in  \mathbb{R}[S_n] \;\; {\rm such \;\;that} \;\; \bm{u} \times \bm{u}_j^*=0\}$. 
Using (\ref{ATstar}) one can then show that the relations (\ref{Tu}) are equivalent to 
\begin{eqnarray}
\forall \vcovect\;, \;\T_{\vcovect} \times \bm{u}^*_j =0 {\rm\;\;for \;\; all\;\;} j=1,\cdots,m.
\end{eqnarray}
This shows that for a tensor obeying the symmetries (\ref{Tu}) all the $\T_{\vcovect}$ belong to the intersection $L$ of the ideals $L_j$. Using the results summarized after Eq.~(\ref{Decirredleft}), one can find a generating idempotent $e$ of this ideal which can be decomposed into primitive pairwise orthogonal idempotents $\bm{e}_k$ such that $\bm{e}= \bm{e}_1 +\cdots \bm{e}_m$. These idempotents decompose every $\T_{\vcovect}$ as
\be
\T_{\vcovect} = \T_{\vcovect} \bm{e} = \T_{\vcovect} \bm{e}_1 + \cdots + \T_{\vcovect} \bm{e}_m.
\ee
This decomposition is equivalent to 
\be
\bm{\T} = \bm{e}_1^* \bm{\T}+ \cdots + \bm{e}_m^* \bm{\T},
\ee
and gives a decomposition of the tensor $\bm{\T}$ into tensors each belonging to a symmetry class \cite{Boerner} generated by the idempotent $\bm{e}_k^*$ and in one to one correspondence with an irreducible representation of $S_n$ \cite{FiedlerHab,Fiedler1}.

\subsection{Tensor product and the Littlewood-Richardson rule} \label{LittlewoodRichardson}
Consider two tensors each belonging to some tensor symmetry class generated by some standard tableaux $\lambda$ and $\mu$ respectively (i.e, speaking more properly, generated by the Young Symmetrizers associated with the corresponding standard Young tableaux). As such, these symmetry classes are irreducible but the tensor product of these two tensors belong to a symmetry class which is in general reducible. The rule to obtain the decomposition of the tensor product into irreducibles is known as the Littlewood-Richardson rule and goes as follows. 
\begin{itemize}
\item Since tensor products are  commutative (and associative) one can choose a convenient order of the tensor product.
Then Label all the boxes of each row of the second tableau with the same letter following some canonical order, e.g., $a,b,c,\dots$, and going from top to bottom.

\item Add the boxes of the second tableau one by one, starting from $a's$ then $b's$. and so on, to the first Young diagram such that (i) the resulting diagram is a valid Young diagram, (ii) no column contain the same label, and (iii) in the resulting diagram, when read from right to left and top to bottom, the number of a's encountered $\geq$ the number of b's encountered $\geq$ $\dots$.
\end{itemize}
Applying this rule, we obtain a set of Young diagrams (filled in part with letters $a,b,c,\dots$) each representing one particular irreducible component in direct sum decomposing the tensor product of the symmetry classes corresponding to the two tensors considered initially. In particular, the multiplicity $m_\rho$ corresponding to the representation associated with a given Young diagram $\rho$ is given by the number of times (counted after the removal of the letters $a,b,c,\dots$) this diagram $\rho$  is created by the application of the Littlewood Richardson rule.

As an example, the tensor product (denoted here by $\otimes$) 

\ba \label{tensorproductexample}
\ytableausetup{mathmode,aligntableaux=top}
\ydiagram {2,1} \hfill \otimes \hfill \ytableaushort {aa,b} 
\ea
is decomposed as follows into irreducibles 
\ba
\ytableausetup{mathmode,aligntableaux=top}
\ytableaushort{\none \none a a, \none b} *{4,2}
\oplus
\ytableaushort{\none\none a a,\none,b} *{4,1,1} 
\oplus 
\ytableaushort{\none \none a, \none a , b} *{3,2,1} 
\oplus 
\ytableaushort{\none \none a,\none b,a} *{3,2,1} 
 \oplus 
\ytableaushort{\none \none a, \none ab} *{3,3} 
\oplus 
\ytableaushort{\none \none a,\none,a,b} *{3,1,1,1} 
\oplus 
\ytableaushort{\none \none,\none a, a b} *{2,2,2} 
\oplus 
\ytableaushort{\none \none, \none a, a ,b} *{2,2,1,1}. \nonumber
\ea
We see in particular that the Young diagram 
\ytableausetup{mathmode,aligntableaux=top}
\ytableaushort{\none \none \none, \none \none, \none} *{3,2,1}~~
appears two times above and hence the corresponding representation has multiplicity two in the tensor product (\ref{tensorproductexample}).

The above described method does work for mere tensor products, however it is not appropriate to decompose plethysms. Indeed, due to the extra symmetry existing for Plethysms as compared to simple tensor products, a Plethysm has in general a decomposition in terms of irreducibles that is smaller than the mere tensor product it is built from. In order to decompose Plethysms, we will use in this work the link between symmetric functions and representations of the symmetric group summarized in the next section.

\section{Symmetric Functions, Representation of the permutation group and Plethysms} \label{AppD}
\subsection{Symmetric and Schur functions} 

We first define an \textbf{\textit{ordered partitions}} of an integer $n$ which, in the context of this work, is a sequence of positive (and, we stress, possibility null, i.e. element of $\mathbb{N}$) numbers  $\alpha = (\alpha_1,\alpha_2, \dots )$ whose sum equals to $n$ and where the order of the numbers $\alpha_k$ matters, i.e. partitions containing the same integers $\{ \alpha_1,\alpha_2, \dots \}$ in a different order are considered as different. It is easy to see that the set of all the types of all semi standard Young tableaux built from a given Young diagram $\mu$ with $n$ boxes filled with integers $\{1,2,3,\cdots\}$ 
is a subset of the set of all ordered partition of $n$ and we shall call this set $\aleph_\mu$. For example, considering the Young diagram 
$\mu = $\,\ydiagram{2,1}~, we get~, filling it in a semi-standard way with entries in the labels $\{1,2,3\}$, the set of ordered partitions 
$\aleph_\mu=\{(2,1,0),(2,0,1),(1,2,0),(1,1,1),(1,0,2),(0,2,1),(0,1,2),(1,1,1)\}$. Notice e.g. that the partition $(3,0,0)$ does not belong to this set but also that the partition $\{1,1,1\}$ does appear twice as it corresponds to the two different standard Young tableaux \ytableausetup{mathmode,aligntableaux=top}\ytableaushort{1 2,3} *{2,1}~ and  \ytableausetup{mathmode,aligntableaux=top}\ytableaushort{1 3,2} *{2,1}~.

A \textbf{\textit{homogeneous symmetric function of degree n}} is a formal power series of a set of variables $x = (x_1,x_2,\dots)$
\be \label{defnsym}
f(x) = \sum_{\alpha} c_{\alpha} x^{\alpha}
\ee
where $\alpha = (\alpha_1,\alpha_2, \dots )$ is an ordered partition of $n$,
$ x^\alpha$ denotes the monomial $x_1^{\alpha_1} x_2^{\alpha_2}\dots $ and the coefficients $c_\alpha$ are chosen such that any permutation of the variables leaves the function invariant. For example consider $n=2$ and $x=(x_1,x_2)$. Then $\alpha$ ranges over the ordered partitions of $2$ given by $(0,2),(2,0),(1,1)$  and we choose $c_{(0,2)} = c_{(2,0)}$ in order to obtain the symmetric function of degree two in two variables given by 
\be 
g(x_1,x_2) = c_{(1,1)} x_1x_2 + c_{(2,0)}\left( x_1^2 + x_2^2\right),  
\ee
A \textbf{\textit{Schur function}} $s_\lambda$ (see e.g. \cite{Stanley}) corresponding to a Young diagram $\lambda$ is a special kind of homogeneous symmetric function defined by 
\be \label{defnschur}
s_{\lambda} = \sum_{\alpha\in \aleph_\lambda} x^{\alpha}
\ee
where the sum ranges over elements of the set $\aleph_\lambda$ which are in one to one correspondence with the semi-standard Young tableaux generated by filling the Young diagram, $\lambda$. For example, when $x=(x_1,x_2,x_3)$ and for the Young diagram $\mu$ defined above  we have the corresponding Schur function
\be 
s_{\mu} = x_1^2 x_2 + x_1^2 x_3 + x_1x_2^2 + x_1x_3^2 + x_2^2 x_3 + x_2x_3^2 + 2x_1x_2x_3. 
\ee
Other examples of Schur functions that will be used here are those corresponding respectively to the Young diagrams $(m)$ and $(1)^m$. Applying the above definition, we get for these two cases
and $p$ variables $x_i$, $1 \leq i \leq p$
\bea
s_{(m)}(x_i) &=& \sum_{1 \leq i_1 \leq i_2 \leq \dots \leq i_m \leq p} x_{i_1} \dots x_{i_m} \label{defschurm},\\
s_{(1)^m}(x_i) &=& \sum_{1 \leq i_1 <i_2 < \dots < i_m \leq p} x_{i_1} \dots x_{i_m} \label{defschur1m},
\eea
where one sees in particular that the Schur function $s_{(1)^m}(x_i)$ is only non trivial if the number of variables $x_i$ is larger than or equal to $m$.  
One can show that the Schur functions form a basis of the homogeneous symmetric functions (see e.g. \cite{Fulton} section 6).

\subsection{Schur functions, representation of the permutation group and tensors} \label{sccorres}
Schur functions provide a powerful way of describing the representations of the permutation group
which can be used to decompose tensor symmetry classes into irreducibles. 

Let us first consider the case of tensor products between elements of the symmetry classes $\Vspace_\lambda$ and $\Vspace_\mu$ each corresponding to some specific Young diagrams $\lambda$ (with $n$ boxes) and $\mu$ (with $m$ boxes). The decomposition of this tensor product into irreducible spaces (corresponding to minimal ideals  irreducible under the action of $S_{n+m}$) is obtained from the Littlewood Richardson rule, as explained above. The use of the Schur functions allows to compute as well this decomposition (including the  multiplicities $m_\rho$  of the irreducible representation characterized by the Young diagram $\rho$ in the tensor product). To do so is enough to consider the ordinary product of Schur functions corresponding to the Young diagrams $\lambda,\mu$, 
\be \label{Schurfuncproduct}
s_\lambda s_\mu = \sum_\rho m_\rho s_\rho,
\ee
from which $m_\rho$ can be read (see e.g. \cite{Fulton} section 5). By the definition of the Schur functions it is obvious that the number of variables in both sides of this equation agrees with the number of labels from which the entries in the semi-standard Young tableau with the shapes $\lambda$ and $\mu$ are chosen. Since all entries in a column in semi-standard tableau must be distinct, a Schur functions $s_{\rho}(x)$ on the right hand side identically vanish if the number of rows in $\rho$ is greater than the number of variables. This implies that the formula (\ref{Schurfuncproduct}), when restricted to a specific number or variables, does not provide a way to calculate the multiplicity $m_{\rho}$ for a particular Young diagram $\rho$ if the number of rows in $\rho$ is greater than the number of variables. The formula (\ref{Schurfuncproduct}) itself is valid for any numbers of variables.

 Schur functions are also very useful to decompose plethysms into irreducibles. Indeed, consider a given plethysm $\mu \circ \lambda $ as introduced in section \ref{Plethysms} (where $\mu \circ \lambda $ just denotes the symmetries of the tensor symmetry class corresponding to $\bm{\T}_{\mu \circ \lambda}$ introduced there), there is a corresponding homogeneous symmetric function denoted $s_{\mu} \circ s_{\lambda}$ and obtained via the so-called characteristic map (see e.g. \cite{McDo} and \cite{Sagan} pp 167-169, 175) denoted $s_{\mu} \circ s_{\lambda}$.
It can be given a combinatorial formulation as follows. We consider $s_{\lambda}$ as defined in \eqref{defnschur}, 
and choose some ordering for the element $\alpha$ of the set $\aleph_\lambda$ entering the sum in \eqref{defnschur}. 
We call $\alpha^{(i)}$ the $i-th$ ordered partition $\alpha$ in this set (note that $\alpha^{(i)}$ is a partition, and hence a sequence of integers $\alpha^{(i)} = (\alpha^{(i)}_1, \alpha^{(i)}_2, \dots )$), and we define the variables $y_i$ as numerous as the $\alpha^{(i)}$. 
Then we considered the Schur function $s_{\mu}$ in the ordered variables $y_{i}$. The plethysm $s_{\mu} \circ s_{\lambda}$ is then just obtained from this function $s_{\mu}$ where $y_i$ is replaced by  $x^{\alpha^{(i)}}$. This yields a symmetric function in the variables $x_j$.
This symmetric function can then be decomposed on the basis of Schur functions as 
\be 
s_{\mu}\circ s_{\lambda}(x) = \sum_{\rho} m_{\rho} s_{\rho}(x). 
\ee
This gives both the decomposition of the Plethysm $\mu \circ \lambda $ into irreducibles corresponding to the Young diagrams $\rho$ which appear in the right hand side of the above, as well as the multiplicity $m_{\rho}$ of the corresponding representation. An identical decomposition with the same multiplicity holds for the decomposition of the tensor symmetry class $\Vspace_{\mu \circ \lambda}$.

In order to illustrate this, let us consider the following plethysm $s_{(2)} \circ s_{(1)^2}(x_1,x_2)$ in two variables $x^1, x^2$.
We have
\be 
s_{(1)^2}(x_1,x_2) = x_1x_2
\ee
taking $y_1 = x_1x_2$ we write $s_{(2)} \circ s_{(1)^2}$ as $s_{(2)}$ in one variable $y_1$
\be 
s_{(2)}(y_1) = y_1^2 = x_1^2 x_2^2 = s_{(2)} \circ s_{(1)^2}(x_1,x_2).
\ee
We further have that 
\be 
s_{(2)} \circ s_{(1)^2} (x_1,x_2) = s_{(2)^2}(x_1,x_2), 
\ee
which shows that the only irreducible inside the plethysm $(2) \circ (1)^2$  is given by one irreducible representation characterized by a specific standard filling of the Young diagram, \ydiagram{2,2}~. This corresponds to the symmetries of the Riemann tensor.


\subsection{Involution on symmetric functions}
In this work we will use the following map inside the set of homogeneous  symmetric functions that will be called the involution map $\Omega$ (as it turns out to be an involution, i.e. it verifies $\Omega^2 = \mathbb{1}$).  Consider a homogeneous  symmetric function 
$f(x)$ which has the following expansion in terms of Schur functions $s_\mu$. Since Schur functions furnish a basis for the  homogeneous  symmetric functions (see e.g. \cite{Fulton} section 6), one can always obtain such a decomposition.
\be 
f(x) = \sum_{\mu} k_\mu s_{\mu}(x), 
\ee 
where $k_\mu$ are real numbers indexed by the Young diagrams $\mu$. 
Then we define $\Omega(f) = g$ as the homogeneous  symmetric function defined by 
\be 
\Omega(f)(x) \equiv g(x) = \sum_{\mu} {k_\mu} s_{\mu^t}(x)
\ee
where we recall (see appendix \ref{AppB}) that $\mu^t$ is the conjugate of the Young diagram $\mu$.  

So defined, the map $\Omega$ is obviously linear and it acts in a non trivial way on plethysms. Indeed, if  $s_{\lambda}, s_{\mu}$ are Schur functions of degree $m,n$ respectively, then (see e.g. \cite{McDo} page 136) 
\be  \label{OMEGAPARITY}
\Omega(s_{\lambda} \circ s_{\mu}) = \begin{cases}
s_{\lambda}\circ \Omega(s_{\mu}) &\text{if $n$ is even,}\\
\Omega(s_{\lambda}) \circ \Omega(s_{\mu})&\text{if $n$ is odd}
\end{cases}
\ee

\end{document}